\documentclass[12pt]{unbthesis}
\usepackage[pdftex]{color, graphicx}
\usepackage{fancyhdr}
\usepackage[english]{babel}
\usepackage{footmisc}
\usepackage{algorithm}
\usepackage{algorithmic}
\usepackage{listings}
\usepackage{fancyvrb}
\usepackage{url}
\title{A Branch and Cut Algorithm for the Halfspace Depth Problem}
\author{Dan Chen}
\predegree{Bachelor of Science, Liaoning University, 2003}
\degree{Master of Computer Science}
\gau{Computer Science}
\supervisor{David Bremner, PhD, Computer Science}
\examboard{Eric Aubanel, PhD, Computer Science, Chair\\ 
 & Patricia Evans, PhD, Computer Science,\\
 & Luis Zuluaga, PhD, Business Administration}
\date{February, 2007}
\copyrightyear{2007}
\newtheorem{theorem}{Theorem}[section]

\DeclareMathOperator{\dist}{dist}

\begin{document}
\unbtitlepage
\doublespacing
\setcounter{secnumdepth}{3} \setcounter{tocdepth}{3}
\pagenumbering{roman} \setcounter{page}{1}
\chapter*{Abstract}
\addcontentsline{toc}{chapter}{Abstract}
The concept of data depth in non-parametric multivariate descriptive statistics is the generalization of the univariate rank method to multivariate data. Halfspace depth is a measure of data depth. Given a set $S$ of points and a point $p$, the halfspace depth (or rank) $k$ of $p$ is defined as the minimum number of points of $S$ contained in any closed halfspace with $p$ on its boundary. Computing halfspace depth is NP-hard, and it is equivalent to the Maximum Feasible Subsystem problem. In this thesis a mixed integer program is formulated with the big-$M$ method for the halfspace depth problem. We suggest a branch and cut algorithm. In this algorithm, Chinneck's heuristic algorithm is used to find an upper bound and a related technique based on sensitivity analysis is used for branching. Irreducible Infeasible Subsystem (IIS) hitting set cuts are applied. We also suggest a binary search algorithm which may be more stable numerically. The algorithms are implemented with the BCP framework from the \textbf{COIN-OR} project.

\singlespace
\renewcommand{\contentsname}{Table of Contents}
\tableofcontents{}
\addcontentsline{toc}{chapter}{Table of Contents}
\listoftables{}
\addcontentsline{toc}{chapter}{List of Tables}
\listoffigures{}
\addcontentsline{toc}{chapter}{List of Figures}
\chapter*{List of Abbreviations}
\addcontentsline{toc}{chapter}{Abbreviations}

\begin{table}[!h]
\begin{tabular}{l l}
ANOVA & Analysis of Variance\\
BCP   & Branch-Cut-Price\\
BIS   & Basic Infeasible Subsystem\\
IIS   & Irreducible Infeasible Subsystem\\
LP    & Linear Programming\\
MAX FS & Maximum Feasible Subsystem\\
MDS   & Minimal Dominating Set\\
MIN IIS COVER & Minimum-Cardinality IIS Set-Covering\\
MIN ULR & Minimum Unsatisfied Linear Relation\\
MIP   & Mixed Integer Program\\
MPS   & Mathematical Programming System\\
NINF  & Number of Infeasibility\\
SINF  & Sum of Infeasibility
\end{tabular}
\end{table}

\doublespace
\pagenumbering{arabic} \setcounter{page}{1}
\chapter{The Halfspace Depth Problem}
\label{chap:pro}

\section{Data Depth}
\label{sec:pro.dd}
\emph{Halfspace depth} is a measure of \emph{data depth}. The term data depth comes from non-parametric multivariate descriptive statistics. Descriptive statistics is used to summarize a collection of data, for example by estimating the center of the data set. In non-parametric statistics, the probability distribution of the population is not considered, and the test statistics are usually based on the rank of the data. In multivariate data analysis, every data item consists of several elements (i.e. is an $n$-tuple). The idea of data depth in multivariate data analysis is to generalize the univariate rank method to tuple data, and order the data in a center-outward fashion. Since the tuple data items can be represented as points in Euclidean space $\mathbb{R}^{d}$, these two terms are used interchangeably in this thesis. The rank or depth of a point measures the centrality of this point with respect to a given set of points in high dimensional space. The data with the highest rank is considered the center or median of the data set, which best describes the data set.

In $\mathbb{R}^{1}$, the median holds the properties of high \emph{breakdown point}, \emph{affine equivariance}, and \emph{monotonicity}. Breakdown point of a measure is the fraction of the input that must be moved to infinity before the median moves to infinity. In $\mathbb{R}^{1}$, the median has a breakdown point of $\frac{1}{2}$~\cite{Aloupis}.  After an affine transformation on the data set, the median will not be changed. Therefore, median is affine equivariant. When extra data is added to one side of the data set, the median tends to move to that side, never moving to the opposite side. This property is called monotonicity. A good measure of data depth should also hold these properties, ideally, to the same degree as median does.

Many measures of data depth have been introduced, such as halfspace depth~\cite{Hodges,Tukey}, \emph{convex hull peeling depth}~\cite{Barnett,Shamos}, \emph{Oja depth}~\cite{Oja}, \emph{simplicial depth}~\cite{Liu}, \emph{majority depth}~\cite{Singh}, \emph{regression depth}~\cite{Rouss}, and so on. The surveys~\cite{Aloupis,Fukuda1,Liu2,Rafalin} give detailed introductions to these measures.

\section{Halfspace Depth}
\label{sec:pro.hd}
The halfspace depth is also called Tukey depth. Given a set $S$ of points and a point $p$ in $\mathbb{R}^{d}$, the halfspace depth of $p$ is defined as the minimum number of points of $S$ contained in any closed halfspace with $p$ on its boundary. The point with the largest depth is called \emph{halfspace median} or \emph{Tukey median}.

\begin{figure}[!ht]
  \centering
  \includegraphics[width=0.5\textwidth]{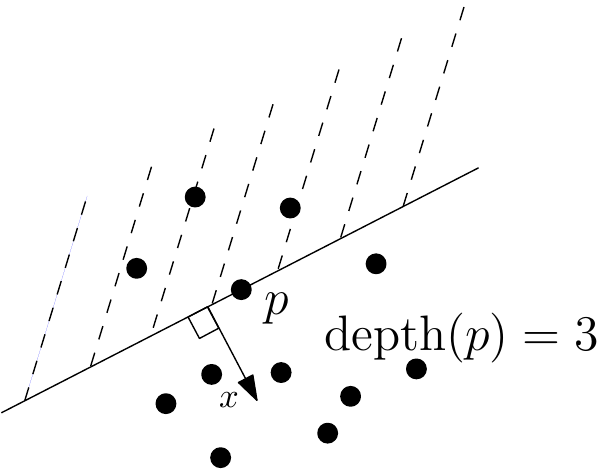}
  \caption{An example of halfspace depth in $\mathbb{R}^{2}$}
  \label{fig:hd1}
\end{figure}

\begin{figure}[!ht]
  \centering
  \includegraphics[width=0.5\textwidth]{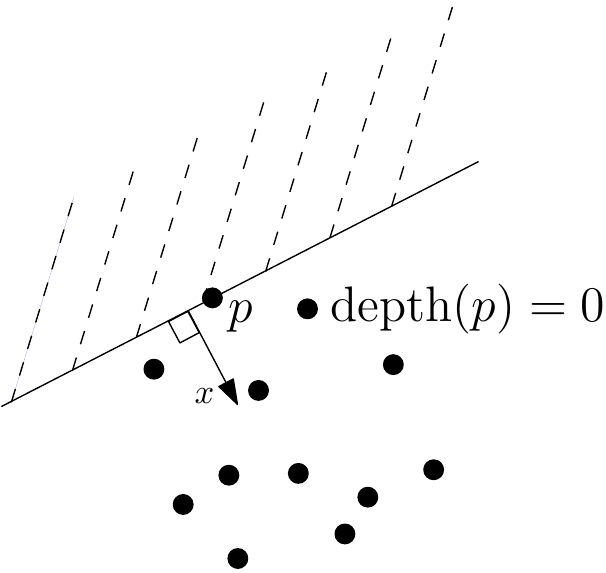}
  \caption{Another example of halfspace depth in $\mathbb{R}^{2}$}
  \label{fig:hd2}
\end{figure}

In Figure~\ref{fig:hd1}, the halfspace depth of $p$ is $3$, because at least three points will be contained by the closed halfspace with $p$ on its boundary. And the depth of point $p$ in Figure~\ref{fig:hd2} is $0$.

The halfspace depth of point $p$ can also be described as:
\begin{equation}
  \label{eq:1.1}
  \min_{x \in \mathbb{R}^{d} \backslash 0} | \{ q \in S  | \langle x , q \rangle \leq \langle x , p \rangle \} |
\end{equation}
where $x$ is the outward normal vector of the closed halfspace. From Figure~\ref{fig:hd1}, we can see that if a point $q$ is contained in the closed halfspace, the corresponding inequality $\langle x , q \rangle \leq \langle x , p \rangle$ will be satisfied. So we are trying to find an $x$ which can minimize the number of satisfied inequalities. Minimizing the number of the points contained in the halfspace is equivalent to maximizing the number of points excluded from the halfspace. Therefore, the definition of halfspace depth can also be described as:
\begin{equation}
  \label{eq:1.2}
  | S | - \max_{x \in \mathbb{R}^{d}} | \{ q \in S  | \langle x , q \rangle > \langle x , p \rangle \} |
\end{equation}
When a point is excluded from the halfspace, the corresponding inequality in \eqref{eq:1.2} is satisfied. Then the problem is to find a vector $x$ that maximizes the number of satisfied inequalities.

A data set is said to be in general position if it has no ties, no more than two points on the same line, no more than three points on the same plane and so forth. If the data set is in general position, computing the halfspace depth of a point is identical to the \emph{open hemisphere problem} introduced by Johnson and Preparata. Given a set of $n$ points on the unit sphere $S^{d}$ in $\mathbb{R}^{d}$, the open hemisphere problem is to find an open hemisphere of $S^{d}$ that contains as many points as possible. This problem is NP-complete if both $n$ and $d$ are parts of the input~\cite{Johnson}.

\subsection{Properties of Halfspace Depth}
\label{sec:pro.hdp}
\begin{figure}[!ht]
  \centering
  \includegraphics[width=0.5\textwidth]{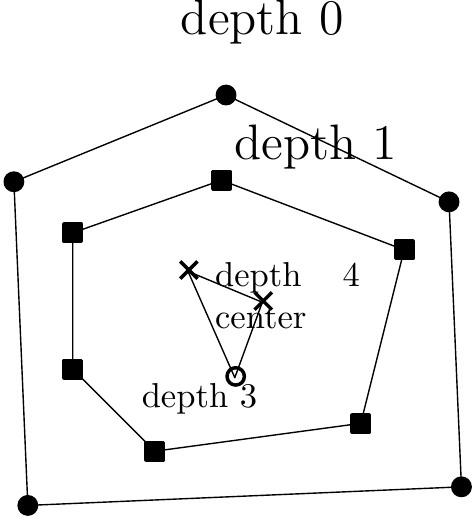}
  \caption{Halfspace depth contours}
  \label{fig:contou}
\end{figure}

If the data set is in general position, the depth of the halfspace median is in the range of $\lceil \frac{n}{d + 1} \rceil - 1$ and $\lceil \frac{n}{2} \rceil - 1$~\cite{Donoho2} (our depth values differ by one from the ones in~\cite{Donoho2}). The halfspace median might not be unique, but the measure of halfspace depth is preferred by statisticians compared with other measures because it has the following properties:
\begin{description}
\item[High Breakdown Point] In $\mathbb{R}^{d}$, the breakdown point halfspace depth is at least $\frac{1}{d + 1}$ and can be as high as $\frac{1}{3}$ when $d$ is greater than $2$~\cite{Donoho2,Donoho1}.
\item[Affine Equivariance] After an affine transformation of the data set, the rank value of any data item will not be changed.
\item[Monotonicity] The halfspace median tends to move to the location where data is added.
\item[Convex and Nested] The boundary of the set of data with depth at least $k$ is called the \emph{contour} of depth $k$~\cite{Donoho2} (see Figure~\ref{fig:contou}). The contours are all convex. The contours are also nested. The contour of depth $k + 1$ is completely contained by the contour of depth $k$~\cite{Donoho2}.
\end{description}

\section{Overview of This Thesis}
\label{sec:pro.over}

In this chapter we introduced the definition of the halfspace depth and some basic properties. In Chapter~\ref{chap:mfs} we show that the halfspace depth problem is equivalent to the \emph{maximum feasible subsystem (MAX FS)} problem. In Chapter~\ref{chap:mip} we discuss different integer problem formulations for the halfspace depth problem. In Chapter~\ref{chap:heur} we introduce the heuristic algorithm developed by Chinneck for the MAX FS problem. In Chapter~\ref{chap:bac} we introduce the branch and cut method for solving general integer programs. In Chapter~\ref{chap:alg} we introduce our branch and cut algorithm for the halfspace depth problem. We also introduced a binary search strategy in this chapter, due to the fact that we can not check the accuracy of the results. In Chapter~\ref{chap:impl} we introduce the details of the implementation of our algorithm, which is implemented with the BCP framework. BCP is also briefly introduced in this chapter. In Chapter~\ref{chap:test} we give some testing results and benchmark the performance of our algorithm. In Chapter~\ref{chap:concl} we summarize the work in this thesis, and give some conclusions.

Throughout this thesis we assume that the reader is familiar with linear programming, integer programming, and combinatorial optimization. For linear programming, we refer to the books by Chv\'atal~\cite{Chvatal}, Hillier and Lieberman~\cite{Hillier}. 

\chapter{Maximum Feasible Subsystem}
\label{chap:mfs}

\section{Introduction}
\label{sec:mfs.1}
The halfspace depth problem has a strong connection with the maximum feasible subsystem problem. If a linear system has no solution, we say this system is \emph{infeasible}. Given an infeasible linear system, the MAX FS problem is to find a maximum cardinality feasible subsystem. This problem is NP-hard~\cite{Chakravarti,Sankaran}, and it is also hard to approximate~\cite{Amaldi}. Pfetsch shows several applications of MAX FS in~\cite{Pfetsch}, for example, linear programming, telecommunications, and machine learning. 

When point $p$ is contained in the convex hull of $S$, and $p$ is on the boundary of a closed halfspace, as shown in Figure~\ref{fig:hd1}, there must be some data contained by the halfspace. Then the set of inequalities
\begin{equation}
  \label{eq:2.1}
  \langle x , q \rangle > \langle x , p \rangle \qquad \forall q \in S
\end{equation}
or
\begin{equation}
  \label{eq:2.2}
  \langle x , q - p \rangle > 0 \qquad \forall q \in S
\end{equation}
in \eqref{eq:1.2} can not be satisfied at the same time, in other words, \eqref{eq:2.2} is an infeasible linear system. To compute the halfspace depth of point $p$ is to find the maximum number of inequalities in \eqref{eq:2.2} that can be satisfied at the same time, or say to find the maximum feasible subsystem of \eqref{eq:2.2}. Therefore, the halfspace depth problem is a MAX FS problem. Of course, if $p$ is outside of the convex hull of $S$, as $s$, \eqref{eq:2.2} will be feasible, and the depth for $p$ will be $0$.

The MAX FS problem can also be seen as finding a minimum cardinality set of constraints, whose removal makes the original infeasible system feasible. This problem is called the \emph{minimum unsatisfied linear relation (MIN ULR)} problem.

\section{Irreducible Infeasible Subsystems}
\label{sec:mfs.2}
\begin{figure}[!ht]
  \centering
  \includegraphics[width=0.33\textwidth]{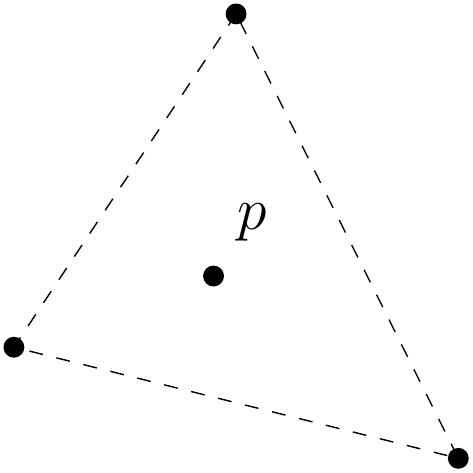}
  \caption{An MDS in $\mathbb{R}^{2}$}
  \label{fig:mds1}
\end{figure}

\begin{figure}[!ht]
  \centering
  \includegraphics[width=0.21\textwidth]{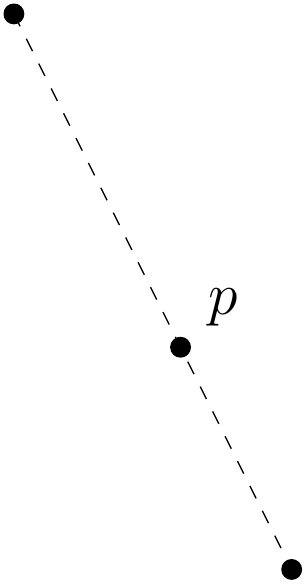}
  \caption{A degenerate MDS in $\mathbb{R}^{2}$}
  \label{fig:mds2}
\end{figure}

In an infeasible linear system, an \emph{irreducible infeasible subsystem (IIS)} is a subset of constraints that itself is infeasible, but any proper subsystem is feasible. If a subset of points $A$ of $S$ forms a simplex which contains $p$, the inequalities in \eqref{eq:2.2} defined by $A$ form an IIS. For example, in Figure~\ref{fig:mds1} three points form a simplex which contains point $p$, so they can not be simultaneously excluded from any closed halfspace with boundary through $p$. Then the corresponding inequalities form an infeasible system. The system is irreducible because if any point is removed, the other two can be excluded at the same time. The point set $A$ is a \emph{minimal dominating set (MDS)}, which is is a set of points forming a minimal convex hull that contains $p$~\cite{David}. A degenerate MDS is shown in Figure~\ref{fig:mds2}, where the three points are collinear.

Every infeasible system contains one or more IISs. To make the original system feasible, we need to delete at least one inequality from every IIS, in other words, we need to delete a hitting set of all IISs in the infeasible system. The \emph{minimum-cardinality IIS set-covering (MIN IIS COVER)} problem is to find the smallest cardinality set of constraints to hit all IISs of the original system (this problem is a \emph{minimum hitting set} problem, although it is called a set cover problem in~\cite{Chinneck, Parker}). The MIN IIS COVER set (hitting set) is the smallest set of constraints whose removal makes the original infeasible system feasible. Hence, the MIN IIS COVER problem is identical to the MIN ULR problem, and hence the MAX FS problem.

Parker gives a method for the MAX FS problem in~\cite{Parker}, and Pfetsch further develops this method in~\cite{Pfetsch}. Due to the fact that the infeasible system could contain an exponential number of IISs with respect to the number of constraints and the number of variables~\cite{Chakravarti}, the main idea of Parker's method is finding a subset of IISs in the whole problem and solving an integer program to find a minimum hitting set in each iteration. If the hitting set hits all IISs in the original infeasible system, the optimum solution is found. If not, find some IISs that are not hit by the current hitting set, then find (with an integer program) a new minimum hitting set that also hits the new IISs.

An important part of this method is finding IISs. Given a linear system $Ax \geq b$, where $A \in \mathbb{R}^{m \times n}$ and $b \in \mathbb{R}^{m}$, the following polyhedron:
\begin{equation}
  \label{eq:2.3}
  P = \{y \in \mathbb{R}^{m} | y^{T}A = 0, y^{T}b = 1, y \geq 0\}
\end{equation}
is defined as the \emph{alternative polyhedron}. Each vertex of $P$ corresponds to an IIS in the original infeasible system~\cite{Gleeson,Khach,Parker,Pfetsch}. More precisely, the set of non-zero supports of a vertex corresponds to an IIS.

In this chapter we introduced the maximum feasible subsystem problem and the irreducible infeasible subsystem, demonstrated that it is equivalent to the halfspace depth problem. In the next chapter we will explore the mixed integer program modeling of the halfspace depth problem.

\chapter{Mixed Integer Program (MIP) Formulation}
\label{chap:mip}
Parker suggests two integer program formulations for the MIN IIS COVER problem in~\cite{Parker}. One is applying the big-$M$ method (see~\cite{Parker} and~\cite{Pfetsch}) to the inequalities in the infeasible system, and the other is based on the IIS inequalities. In this chapter, we will introduce these MIP formulations.

\section{The Infeasible System}
\label{sec:mip.1}
Suppose we have a group of data $\{A_{1}, A_{2}, \ldots, A_{n}\}$ and a point $A_{p}$ in Euclidean space $\mathbb{R}^{d}$, and $x$ is the normal vector of the halfspace that defines the halfspace depth of $A_{p}$. Finding the halfspace depth of $A_{p}$ is equivalent to finding the MIN IIS COVER $\Gamma$ of the following system:
\begin{eqnarray}
  \label{eq:3.1}
  \sum_{i = 1}^{d} (A_{1}^{i} - A_{p}^{i}) x_{i} & > & 0 \nonumber \\
  \sum_{i = 1}^{d} (A_{2}^{i} - A_{p}^{i}) x_{i} & > & 0 \\
  \vdots  & & \vdots \nonumber \\
  \sum_{i = 1}^{d} (A_{n}^{i} - A_{p}^{i}) x_{i} & > & 0 \nonumber
\end{eqnarray}
The depth of $A_{p}$ is $|\Gamma|$.

\section{Parker's Formulation}
\label{sec:mip.parker}
Parker reports that the integer program formulated with the big-$M$ method is hard to solve when the problem size is large. In~\cite{Parker}, Parker deals with the MIN IIS COVER problem with an integer program formulated using the IIS inequalities. First of all, let us introduce the IIS inequalities. MIN IIS COVER is a minimum hitting set problem, and the hitting set has at least one constraint in common with every IIS in the infeasible system. For an IIS $C$ in \eqref{eq:3.1}, we can use the binary variables associated with the constraints in $C$ to formulate an inequality like
\begin{equation}
  \label{eq:mip.iisinq}
  \sum_{t \in C} s_{t} \geq 1
\end{equation}
where $s_{t}$ is the binary variable associated with constraint $t$ in \eqref{eq:3.1}.

Using the IIS inequalities, a hitting set integer program is formulated in the following form:
\begin{eqnarray}
  \label{eq:mip.iismip}
  \textrm{minimize} \qquad \sum_{i = 1}^{n} s_{i} & & \nonumber \\
  \textrm{subject to} \qquad
  \sum_{i \in C} s_{i} & \geq & 1 \qquad \forall C \quad \textrm{(IIS of system~\ref{eq:3.1})} \\
  s_{i} & \in & \{0 , 1\} \qquad \forall i \in \{1, 2, \ldots, n\} \nonumber
\end{eqnarray}

As we mentioned in Section~\ref{sec:mfs.2}, Parker's strategy is to first find a small set of IISs and formulate an integer program (a sub-program of \eqref{eq:mip.iismip}). After obtaining the optimum solution to the initial integer program, find some IISs that are not hit by the solution, add the corresponding IIS inequalities into the integer program and resolve it. The process stops when the solution hits all IISs in the infeasible system.

\section{An MIP with the Big-$M$ Method}
\label{sec:mip.3}
Instead of using the hitting set integer program, we treat the halfspace depth problem with the big-$M$ method. To formulate an integer program for the halfspace depth problem, the strict inequalities in system \eqref{eq:3.1} need to be transformed into non-strict ones. From \eqref{eq:3.1}, we can derive the following possibly infeasible system:
\begin{eqnarray}
  \label{eq:3.2}
  \sum_{i = 1}^{d} (A_{1}^{i} - A_{p}^{i}) x_{i} & \geq & \epsilon \nonumber \\
  \sum_{i = 1}^{d} (A_{2}^{i} - A_{p}^{i}) x_{i} & \geq & \epsilon \\
  \vdots  & & \vdots \nonumber \\
  \sum_{i = 1}^{d} (A_{n}^{i} - A_{p}^{i}) x_{i} & \geq & \epsilon \nonumber
\end{eqnarray}
where $\epsilon$ is a small positive real number. We can get rid of the $\epsilon$ in \eqref{eq:3.2} by dividing both sides of the inequalities by the $\epsilon$. Then we will have the following system:
\begin{eqnarray}
  \label{eq:3.10}
  \sum_{i = 1}^{d} (A_{1}^{i} - A_{p}^{i}) x_{i} & \geq & 1 \nonumber \\
  \sum_{i = 1}^{d} (A_{2}^{i} - A_{p}^{i}) x_{i} & \geq & 1 \\
  \vdots  & & \vdots \nonumber \\
  \sum_{i = 1}^{d} (A_{n}^{i} - A_{p}^{i}) x_{i} & \geq & 1 \nonumber
\end{eqnarray}
Because the elements of $x$ are variables, the elements of $\frac{x}{\epsilon}$ are still variables. Hence, the left hand sides of \eqref{eq:3.10} are the same as \eqref{eq:3.2}. For the half\-space depth problem, we formulate a mixed integer program with the big-$M$ method as follows:
\begin{eqnarray}
  \label{eq:3.11}
  \textrm{minimize} \qquad \sum_{j = 1}^{n} s_{j} & & \nonumber \\
  \textrm{subject to} \qquad
  \sum_{i = 1}^{d} (A_{1}^{i} - A_{p}^{i}) x_{i} + s_{1}M & \geq & 1 \nonumber \\
  \sum_{i = 1}^{d} (A_{2}^{i} - A_{p}^{i}) x_{i} + s_{2}M & \geq & 1 \\
  \vdots  & & \vdots \nonumber \\
  \sum_{i = 1}^{d} (A_{n}^{i} - A_{p}^{i}) x_{i} + s_{n}M & \geq & 1 \nonumber \\
  s_{j} & \in & \{0 , 1\} \qquad \forall j \in \{1, 2, \ldots, n\} \nonumber \\
  - \infty \leq & x_{i} & \leq + \infty \qquad \forall i \in \{1, 2, \ldots, d\} \nonumber
\end{eqnarray}

Fixing the binary variable $s_{j}$ to $1$ has the effect of removing constraint $j$ from \eqref{eq:3.1}. The objective function is to minimize the number of constraints that have to be removed for finding a feasible subsystem of \eqref{eq:3.1}. For the general MIN IIS COVER problems, the big-$M$ method may not be practical. As Parker and Pfetsch mentioned the big-$M$ should be big enough to make the infeasible system feasible, but if it is too big, it will bring numerical problems (see~\cite{Parker} for details). On the other hand Pfetsch notes that this method works reasonably well in the digital broadcasting application~\cite{Rossi}. In this thesis we investigate the big-$M$ method for the halfspace method.

In this formulation, it is easy to find a value for $M$ to make \eqref{eq:3.11} feasible, but the value of $M$ should be large enough to guarantee an accurate result. It is easy to see that if $M$ is assigned to $1$, \eqref{eq:3.11} will be feasible, but the optimal solution will not be the MIN IIS COVER of \eqref{eq:3.10} because all the binary variables will be forced to $1$. Let $X_{o}$ be the value of $x$ in the optimum solution of \eqref{eq:3.11}, and which is a MIN IIS COVER of \eqref{eq:3.10}. For some point $A_{t}$ in the input data set, $\langle X_{o} , A_{t} \rangle$ could be a large negative number, which would require the value of $M$ to be very large.

\section{An Alternative MIP}
\label{sec:mip.2}
Because of the difficulty of finding a proper value for $M$ in the big-$M$ method, we can keep the $\epsilon$. Using the big-$M$ method directly on \eqref{eq:3.2}, we can formulate the following mixed integer program:
\begin{eqnarray}
  \label{eq:3.3}
  \textrm{minimize} \qquad \sum_{j = 1}^{n} s_{j} & & \nonumber \\
  \textrm{subject to} \qquad
  \sum_{i = 1}^{d} (A_{1}^{i} - A_{p}^{i}) x_{i} + s_{1}M & \geq & \epsilon \nonumber \\
  \sum_{i = 1}^{d} (A_{2}^{i} - A_{p}^{i}) x_{i} + s_{2}M & \geq & \epsilon \\
  \vdots  & & \vdots \nonumber \\
  \sum_{i = 1}^{d} (A_{n}^{i} - A_{p}^{i}) x_{i} + s_{n}M & \geq & \epsilon \nonumber \\
  s_{j} & \in & \{0 , 1\} \qquad \forall j \in \{1, 2, \ldots, n\} \nonumber \\
  - \infty \leq & x_{i} & \leq + \infty \qquad \forall i \in \{1, 2, \ldots, d\} \nonumber
\end{eqnarray}
In this formulation, the $\epsilon$ should be small enough to guarantee an accurate result of halfspace depth. At the same time, the $M$ should also be big enough. From the definition of \emph{inner product}, we can get the following observation:
\begin{equation}
  \label{eq:3.6}
   x \cdot q = \lVert x \rVert \cdot \lVert q \rVert \cdot \cos\alpha \leq \lVert x \rVert \cdot \lVert q \rVert
\end{equation}
Now we can give a bound $-c \leq x_{i} \leq c$ for each element of vector $x$, where $c$ is a constant number. Suppose point $q_{max}$ is the point with the largest norm value in the data set. Then the $M$ can be set to the value of $\sqrt{d \cdot c^{2}} \cdot \lVert q_{max} \rVert$.

\begin{figure}[!ht]
  \centering
  \includegraphics[width=0.6\textwidth]{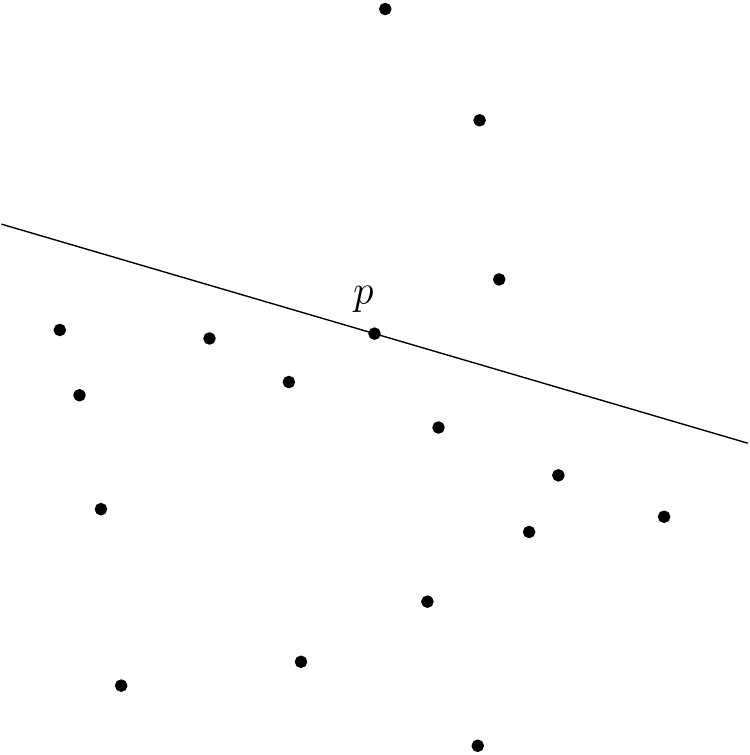}
  \caption{Halfspace depth defined by an open halfspace}
  \label{fig:3.1}
\end{figure}

\begin{figure}[!ht]
  \centering
  \includegraphics[width=0.6\textwidth]{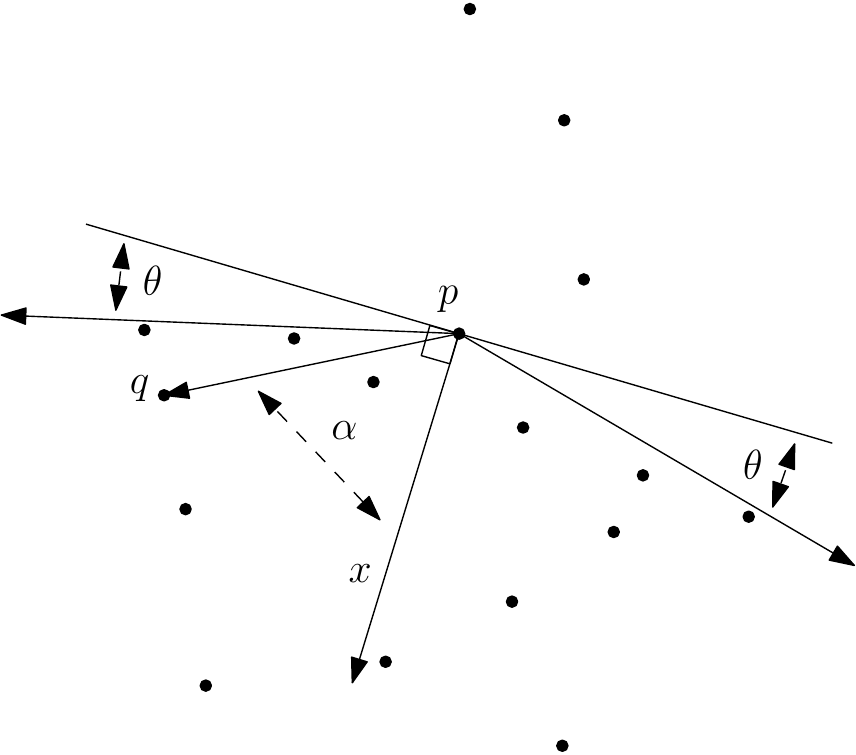}
  \caption{Halfspace depth defined by a cone}
  \label{fig:3.2}
\end{figure}

When computing the halfspace depth, maximizing the number of points contained in an open halfspace is the same as maximizing the number of points contained in a cone. For instance, if the open halfspace in Figure~\ref{fig:3.1} is replaced by a cone (see Figure~\ref{fig:3.2}), we can still have the same depth value for point $p$.

Suppose the angle between the boundary of the cone and the halfspace is $\theta$. Without loss of generality, we can take $p$ as the origin of the space. Let $x$ be an outward normal vector of the halfspace, $q$ be a point contained by the cone, and $\alpha$ be the angle between $x$ and $q$ (which is a point in the data set, thus, it is a vector in the space). The definition of the inner product tells us
\begin{equation}
  \label{eq:3.4}
  x \cdot q = \lVert x \rVert \cdot \lVert q \rVert \cdot \cos\alpha = \lVert x \rVert \cdot \lVert q \rVert \cdot \sin(\frac{\pi}{2} - \alpha) \geq \lVert x \rVert \cdot \lVert q \rVert \cdot \sin\theta
\end{equation}
where the last inequality is illustrated in Figure~\ref{fig:3.2}. Suppose point $q_{min}$ is the point with the smallest norm value in the data set; then, no matter what direction $x$ points to in the optimal solution of \eqref{eq:3.3}, there is always an $x$ which makes the following inequality satisfied for any point $q$.
\begin{equation}
  \label{eq:3.5}
  x \cdot q \geq \lVert x \rVert \cdot \lVert q_{min} \rVert \cdot \sin\theta
\end{equation}

Now the question is how to find the $\theta$ when deciding the $\epsilon$ for \eqref{eq:3.3}. With a proper $\theta$, $\epsilon$ can be set to the value of $\sqrt{d \cdot c^{2}} \cdot \lVert q_{min} \rVert \cdot \sin\theta$.

\begin{figure}[!ht]
  \centering
  \includegraphics[width=0.6\textwidth]{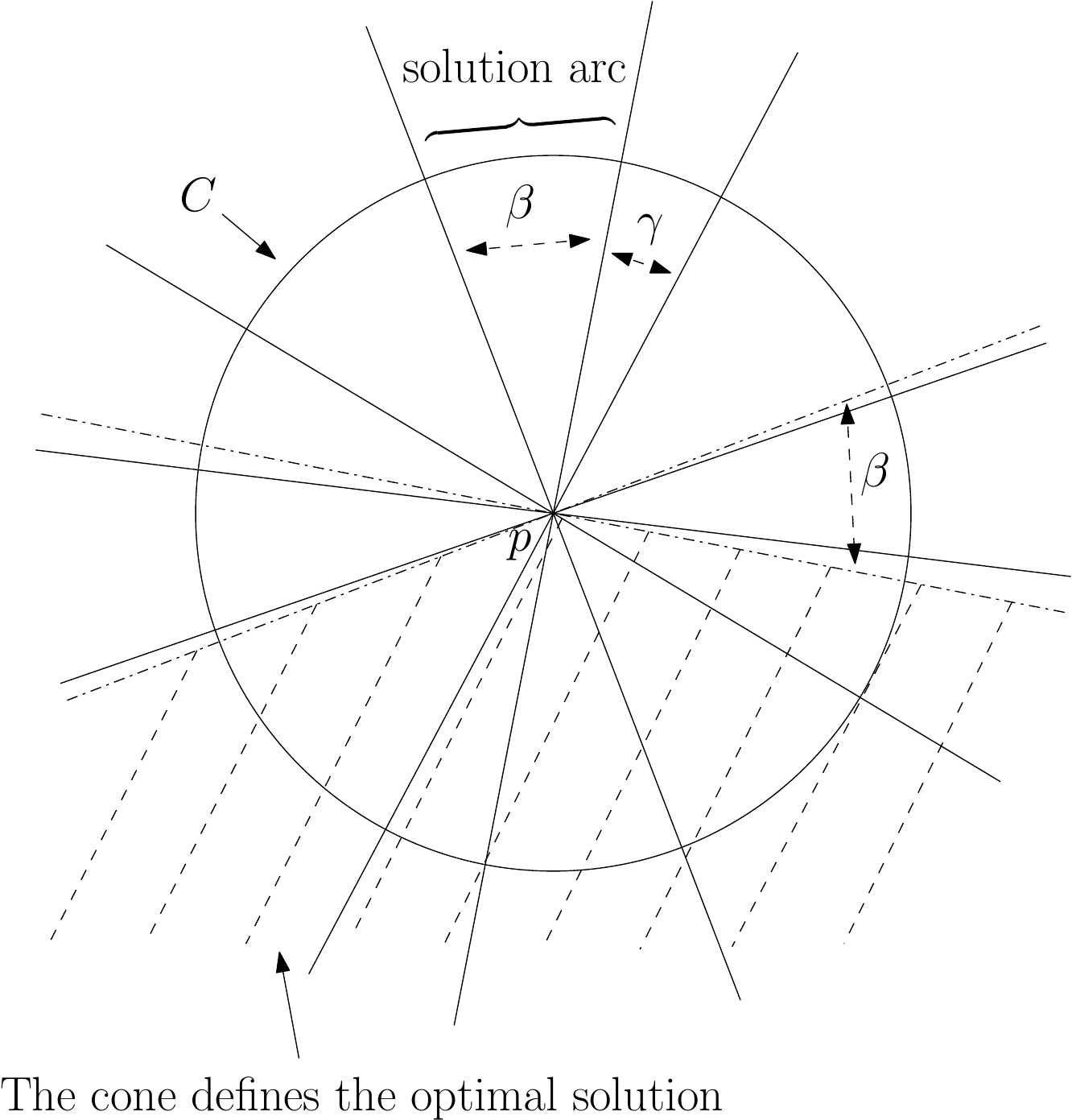}
  \caption{Intersections of the arcs}
  \label{fig:arcs}
\end{figure}

In $\mathbb{R}^{2}$, let $C$ be a circle centered on $p$. A point $q$ defines an arc on $C$ such that every point on this arc defines an $x$ which corresponds to an open halfspace that contains $q$, where $x$ is the outward normal vector of the corresponding halfspace. The arcs are actually half circles. The halfspace depth problem then can be viewed as finding a point which is contained in the largest number of arcs~\cite{Bremner1}. The optimal solutions will form an arc (solution arc) intersected by the largest number of half circles. From Figure~\ref{fig:arcs} we can see that the points excluded from the solution halfspace must be contained in the cone. Suppose the angle corresponding the solution arc is $\beta$, then $\beta = 2\theta$. Therefore, the smallest angle $\gamma$ between the lines gives a lower bound of $2\theta$.

In $\mathbb{R}^{3}$, we can replace the half circles with half spheres. Then the halfspace depth problem can be viewed as finding a point contained in the largest number of spheres. The smallest inscribed cone with $p$ as its apex in the intersections of the half balls that correspond to the spheres will define the lower bound of $\theta$. The lower bound is half of the \emph{opening angle} of the cone. In higher dimensional space the situation will become more complex. We have not got a good idea so far for computing the lower bound.

The value of $\epsilon$ can also be bounded by the method suggested by David Bremner and Achill Sch\"urmann. In this new method all the data are considered as integral data (fractional data can be scaled up to integral data). Then the input data set $S$ will be a subset of the integer lattice~\cite{Gruber}. The following theorem is given by Achill Sch\"urmann.

\begin{theorem}
  Suppose points $\{X_{1}, X_{2}, \ldots, X_{d}\}$ are affinely independent in $\mathbb{R}^{d}$. For any point $X_{i}$ ($X_{i} \in C_{m} := \{ X \in \mathbb{R}^{d} : \lvert X^{j} \rvert \leq m , j = 1,2, \ldots, d\}$). Let $H$ be an affine combination of $\{X_{1}, X_{2}, \ldots, X_{d}\}$, and $H$ does not contain the origin $O$. Then we can have the following statement for the distance from $O$ to $H$, 
\begin{equation}
  \label{eq:3.dist}
   \dist{(H, 0)} \geq (2m\sqrt{d})^{-(d-1)}
\end{equation}
\end{theorem}

\begin{proof}
  Let $l := $
  \begin{displaymath}
    X_{1} + \mathbb{Z}(X_{2} - X_{1}) + \ldots +  \mathbb{Z}(X_{d} - X_{1})
  \end{displaymath}
be a lattice of $\mathbb{Z}^{d}$ within $H$. Let $l_{0} := H \cap \mathbb{Z}^{d} $, then we have $l \subseteq l_{0}$. The distance $h$ of $H$ to a parallel plane containing lattice points is $\frac{\det{\mathbb{Z}^{d}}}{\det{l_{0}}} = \frac{1}{\det{l_{0}}}$. Then $\dist{(H, 0)} \geq h \geq \frac{1}{\det{l_{0}}}$. Since $\frac{\det{l}}{\det{l_{0}}} \in \mathbb{N}$, we have $\det{l} \geq \det{l_{0}}$. Therefore, we have $\dist{(H, 0)} \geq \frac{1}{\det{l}}$. According to Hadamard's inequality, we have $\det{l} \leq \prod_{i=2}^{d} \lVert X_{i} - X_{1} \rVert$. Since $\lVert X_{i} - X_{1} \rVert \leq \textrm{diameter}(C_{m}) = 2m\sqrt{d}$, therefore, $\dist{(H, 0)} \geq (2m\sqrt{d})^{-(d-1)}$.
\end{proof}

\begin{figure}[!ht]
  \centering
  \includegraphics[width=0.75\textwidth]{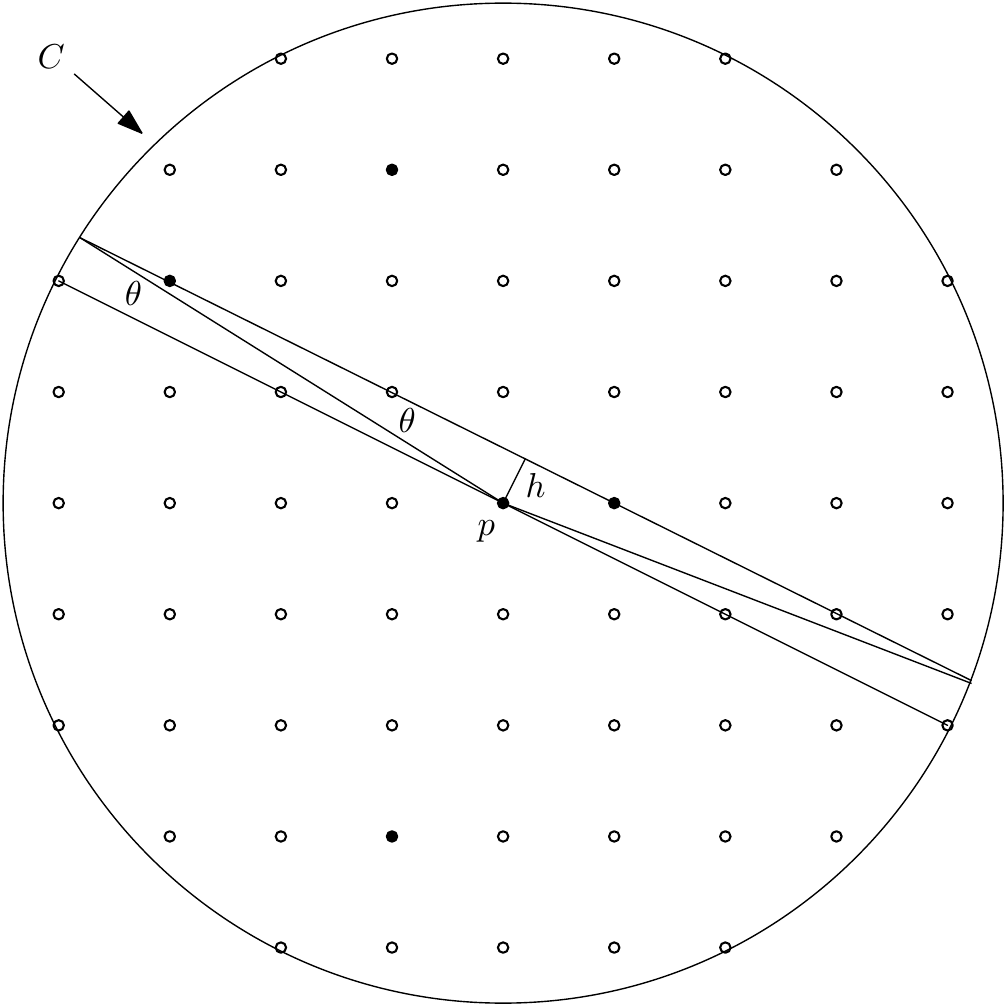}
  \caption{Lattice}
  \label{fig:lattice}
\end{figure}

Let us now return back to the idea of halfspace depth defined by a cone (see~\eqref{eq:3.3}). As shown in Figure~\ref{fig:lattice}, a distance $h$ defines a cone. Point $p$ corresponds to the origin $O$ in the former paragraph. The value of $\sin\theta$ will be $\frac{h}{\textrm{radius}(C)}$.

When the dimension is high, such as $20$, the value of $\epsilon$ based on this lattice idea would be too small to be useful in practice. In our testing, we just set $\epsilon$ to a very small value. If $\epsilon$ is not small enough, it will have the same effect as $M$ not being big enough. Unfortunately, we did not find a way to test whether a solution is accurate.

In this chapter we introduced Parker's MIP formulations and formulated an MIP with the big-$M$ method for the halfspace depth problem. In the next chapter we will introduce a heuristic algorithm for the MAX FS problem, which can be used for the halfspace depth problem.

\chapter{A Heuristic Algorithm}
\label{chap:heur}

\section{Elastic Programming}
\label{sec:heur.1}
Chinneck~\cite{Chinneck2, Chinneck} suggests a heuristic algorithm for the MIN IIS COVER problem. As discussed in Chapter~\ref{chap:mfs}, this is also an algorithm for the halfspace depth problem. This algorithm is based on several observations of elastic programming (a method to solve an integer program~\cite{Brown} according to Chinneck). In elastic programming, every constraint is elasticized by adding a non-negative elastic variable. Chinneck gives the following rules:
\begin{eqnarray}
  \sum_{j}a_{ij}x_{j} \geq b_{i} & \quad   \Longrightarrow   & \quad  \sum_{j}a_{ij}x_{j} + e_{i} \geq b_{i} \nonumber \\
  \sum_{j}a_{ij}x_{j} \leq b_{i} & \quad   \Longrightarrow   & \quad  \sum_{j}a_{ij}x_{j} - e_{i} \leq b_{i} \nonumber \\
  \sum_{j}a_{ij}x_{j} = b_{i}    & \quad   \Longrightarrow   & \quad  \sum_{j}a_{ij}x_{j} + e_{i}' - e_{i}'' = b_{i}\nonumber
\end{eqnarray}
In fully elastic programming the bounds of the variables are also elasticized in following ways:
\begin{eqnarray}
  x_{j} \geq l_{j} & \quad   \Longrightarrow   & \quad x_{j} + \textrm{\b{$e$}}_{j} \geq l_{j} \nonumber \\
  x_{j} \leq u_{j} & \quad   \Longrightarrow   & \quad x_{j} - \textrm{\b{$e$}}_{j} \leq u_{j} \nonumber
\end{eqnarray}
The elastic objective function is to minimize the sum of the elastic variables, which is similar to phase~1 of the \emph{two phase} simplex method~\cite{Chvatal}. After elasticizing, the original infeasible system becomes feasible, and the optimum solution will give some information about the infeasibility in the original system. This elastic programming is also similar to the big-$M$ method. In the big-$M$ method, a set of binary variables with a large coefficient are used to make the infeasible system feasible. When the optimum point of the elastic program is reached, the optimum value of the objective function is called the \emph{sum of the infeasibility (SINF)}. A nonzero elastic variable indicates a violated constraint in the original model, and the number of the nonzero variables is called the \emph{number of infeasibility (NINF)}. As Chinneck observed, the MIN IIS COVER problem is the problem to minimize NINF. At the optimum point, the value of an elastic variable is called the \emph{constraint violation} of the corresponding constraint in the original model. The \emph{reduced cost} of the slack or surplus variable is called the \emph{constraint sensitivity} of the corresponding constraint, which, in fact, is the \emph{shadow price} of the corresponding constraint. The shadow price of a constraint indicates how much the objective value of the optimum solution will be changed by changing the right hand side of the constraint by one unit. For more details about elastic programming, please refer to~\cite{Chinneck1}.

\section{The Heuristic Algorithm}
\label{sec:heur.2}
The heuristic algorithm developed by Chinneck in~\cite{Chinneck2} is based on the following four observations of elastic programming.
\begin{description}
\item[Observation 1] When the elastic program terminates, the constraints associated with non-zero elastic variables form an IIS hitting set.
\item[Observation 2] When the elastic program terminates, if NINF is $1$, the constraint with a non-zero elastic variable forms the MIN IIS COVER.
\item[Observation 3] The SINF will be reduced more by eliminating a constraint in the MIN IIS COVER.
\item[Observation 4] Removing a constraint to which the objective function does not sensitive will not reduce SINF.
\end{description}
Detailed explanations of these observations are available in~\cite{Chinneck2,Chinneck1}. Based on these observations, a heuristic algorithm is given in~\cite{Chinneck1} as follows:
\begin{enumerate}
\item [Step 1] Solve an elastic program of the original infeasible system. If the NINF is $1$, the hitting set is found due to Observation~2. If the NINF is greater than $1$, select the set of constraints with non-zero constraint violation as candidate constraints.
\item [Step 2] For each of these candidate constraints, delete it temporarily and resolve the elastic program and record the corresponding SINF and NINF for this constraint.
\item [Step 3] The constraint with the minimum SINF is a member of the output IIS COVER. Delete this constraint permanently. If the corresponding NINF of this constraint is $1$, the violated constraint is also a member of the output IIS COVER, and the algorithm terminates. If the NINF is greater than $1$, select candidate constraints with the criteria in Step~1, and go to Step~2.
\end{enumerate}

This heuristic may be slow especially when the problem size is big, because in each step we need to solve a linear program for each candidate constraint. Chinneck revised this heuristic in~\cite{Chinneck} to speed up the algorithm. The revision is based on the following two observations:
\begin{description}
\item[Observation 5] For a constraint with constraint violation in the original model, the relative size of the drop in SINF can be estimated by (constraint violation) $\times$ $|$(constraint sensitivity)$|$.
\item[Observation 6] For an constraint with zero constraint violation, the relative size of the drop can be estimated by $|$(constraint sensitivity)$|$.
\end{description}
Based on these two observations, Chinneck gives a new criteria for selecting candidate constraints for Step~2. In the new criteria, the constraints with constraint violation are sorted according to the value of (constraint violation) $\times$ $|$(constraint sensitivity)$|$ in decreasing order, and the first $k$ constraints are selected as candidate constraints; the constraints with zero constraint violation are sorted according to value of $|$(constraint sensitivity)$|$, and the first $k$ constraints are also used as candidate constraints. In practice $k$ can be set to $1$. With the new criteria, fewer candidate constraints will be considered, so the algorithm will be faster although it could be less accurate (neither version has accuracy guarantees). 

In this chapter we introduced elastic programming and Chinneck's heuristic algorithm for the MAX FS problem. In the next chapter we will introduce the branch and cut method for solving general mixed integer programs.

\chapter{The Branch and Cut Paradigm}
\label{chap:bac}

\section{The Branch and Bound Method}
\label{sec:bac.1}
\subsection{Introduction}
\label{sec:bac.1.1}
The \emph{branch and bound} method is an approach for solving discrete and combinatorial optimization problems. Many of these problems can be modeled as integer linear programming problems. An integer linear programming problem is defined by a linear objective function and a set of constraints (linear equalities or inequalities). In addition, some or all variables are restricted to integer values. Correspondingly, the problems are called pure or mixed integer linear programming problems. Any solution that satisfies all these constraints is called a \emph{feasible solution}. The one that maximizes or minimizes the objective function is called the \emph{optimum solution}. To find the optimum solution, all the feasible solutions need to be enumerated and compared, because there are no better ways known for checking whether a given feasible solution is optimum. Unfortunately, the time complexity of the enumerating algorithm will grow exponentially as the problem size increases. Therefore, it is not practical to enumerate all the feasible solutions when the problem is large. The branch and bound method was developed to reduce the number of feasible solutions to test. To simplify this discussion, we explain the branch and bound method on pure integer linear programming problems. It will be obvious at the end that the branch and bound method will work on mixed integer linear programming problems too.

The branch and bound method was developed independently by A.H. Land and A.G. Doig in 1960 and by K.G. Murty, C. Karel, and J.D.C. Little in 1962~\cite{Murty}. The branch and bound method is a \emph{divide and conquer} paradigm. If the original problem is too hard to solve directly, we divide the problem into smaller size subproblems. If any subproblem is still too hard to solve, we will further divide the problem until we can solve them. The branch and bound method manages a problem tree. The original problem is the root of this tree, and the children of a node are the subproblems of the problem associated with the node. This problem tree is called the \emph{search tree}.

\subsection{The Branch and Bound Method}
\label{sec:bac.1.2}
The basic idea of the branch and bound method involves:
\begin{description}
\item[Branching] Choosing and breaking a problem into some small subproblems.
\item[Bounding] Computing the lower (or upper) bounds of the subproblems.
\item[Pruning] Eliminating those subproblems which are not needed for further consideration due to the bounds.
\end{description}
The branch and bound method gives us a way of enumerating the feasible solutions implicitly \textit{i.e.}, by partial enumeration. At any point of the optimization process, the status of the algorithm is defined by the current best feasible solution and the unexplored space of the feasible solutions. We assume the original problem is a minimization problem (a maximization problem can be easily transformed into a minimization problem). The following gives a detailed description about the three basic steps.

\subsubsection*{Branching}
In the branching step, some additional constraints are added into the original set of constraints. With the new constraints, we get a set of new subproblems which are called \emph{candidate problems} for further consideration. Every candidate problem has a (possibly empty) set of feasible solutions. In other words, the set of original feasible solutions is divided into disjoint subsets (whose union is the original set of feasible solutions).
\begin{figure}[!ht]
  \centering
  \includegraphics[width=0.5\textwidth]{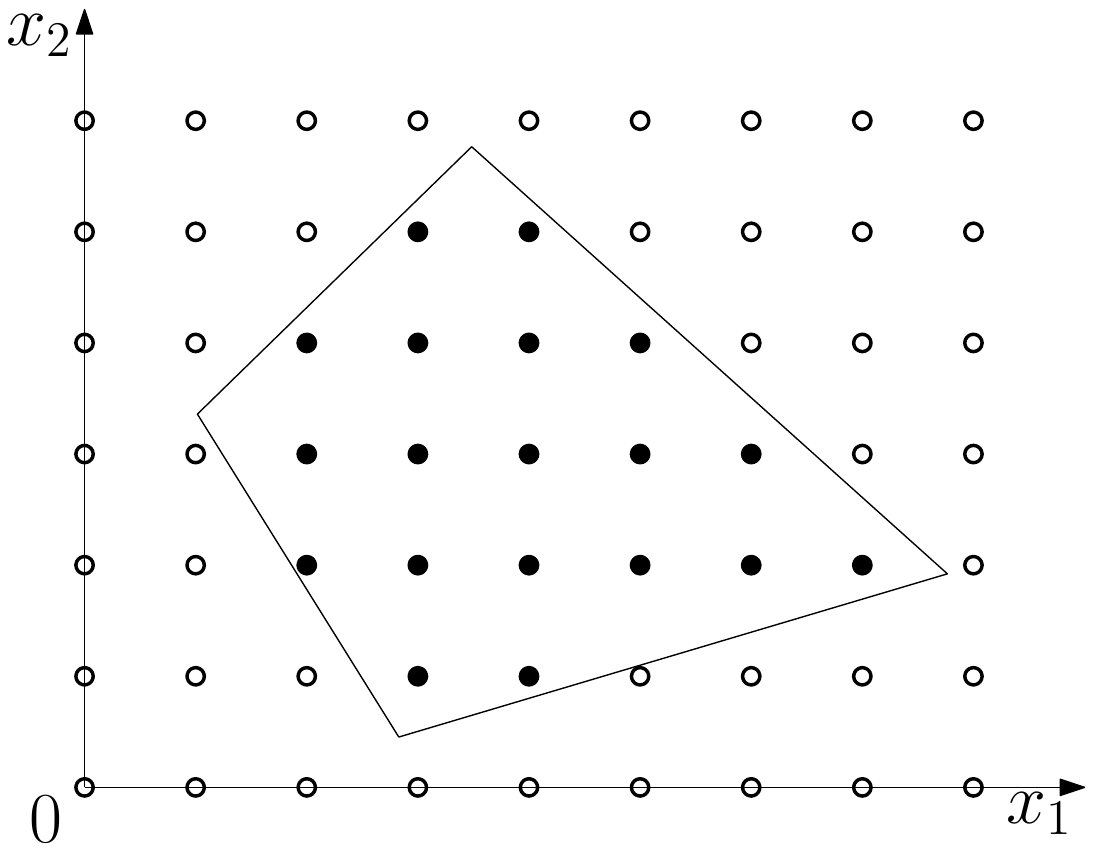}
  \caption{Before branching}
  \label{fig:5.1}
\end{figure}

\begin{figure}[!ht]
  \centering
  \includegraphics[width=0.5\textwidth]{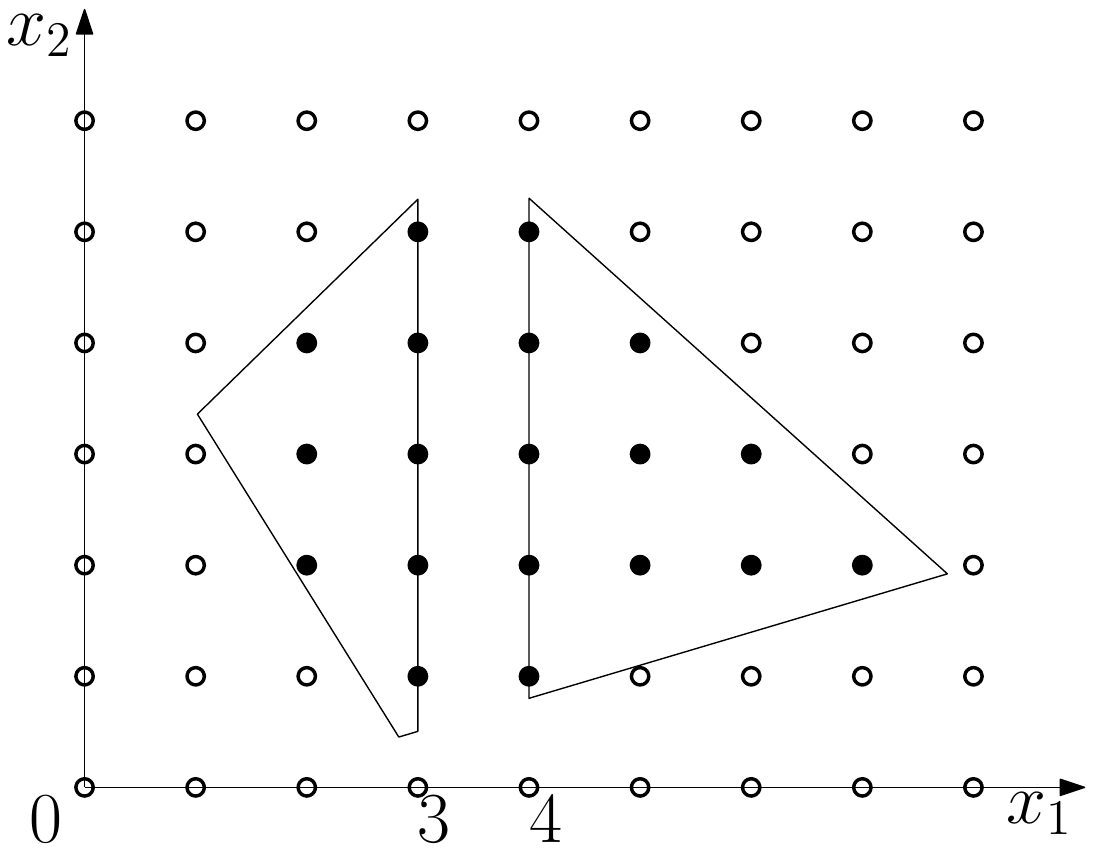}
  \caption{After branching}
  \label{fig:5.2}
\end{figure}

Suppose the feasible region is defined by the polyhedron in Figure~\ref{fig:5.1}. After adding $x \leq 3$ to original problem, we get the subproblem on the left hand side in Figure~\ref{fig:5.2}. After adding $x \geq 4$, we get the subproblem on the right hand side in Figure~\ref{fig:5.2}. Notice that any integer solution must satisfy one of these constraints. The variable $x$ is called the \emph{branching variable}.

\subsubsection*{Bounding}
In the bounding step, for a given subproblem, both the lower and upper bounds of the optimum \emph{objective value} (the result of the objective function) are computed and used. The upper bound bounds the optimum objective value from above, which means that the optimum objective value will not be greater than the upper bound. The upper bound is a global bound, and it bounds every subproblem in the search tree. Therefore, only one upper bound is kept in the whole optimization process. When a smaller upper bound is found, the upper bound for the algorithm will be updated to this smaller value. The solution for the current upper bound is called the current \emph{incumbent} which is the current best feasible solution of the problem. We can simply set the upper bound to positive infinity at the beginning of the algorithm, and when the optimum objective value of a candidate problem is found, updating the upper bound and the incumbent. Another strategy is finding a feasible solution with a heuristic algorithm at the beginning, and setting the upper bound to the objective value of this solution.

On the other side, the lower bound bounds the optimum objective value from below, which means that the optimum objective value will not be smaller than the lower bound. Every candidate problem has its own lower bound which is no larger than any objective value of its feasible solutions. The lower bound is a local bound, and every subproblem has its own lower bound. When computing the lower bound, we hope that the lower bound is as close to the optimum objective value as possible, and that we spend as little effort as possible~\cite{Murty}. One strategy for computing the lower bound is solving a relaxed problem. We can simply remove or relax some constraints of the original hard problem, and then we get a relaxed problem which can be solved with an efficient algorithm. Because the relaxed problem has fewer or looser constraints than the original problem, the set of feasible solutions for the original problem is a subset of the set of feasible solutions of the relaxed problem. Thus the minimum objective value of the relaxed problem must be smaller than or equal to that of the original problem. Therefore, the minimum objective value of the relaxed problem is usually used as the lower bound of the original problem. \emph{Linear programming (LP) relaxation} is the most widely used relaxation. In the LP relaxation, the constraints that restrict variables to integer values are removed.

\subsubsection*{Pruning}
\begin{figure}[!ht]
  \centering
  \includegraphics[width=0.5\textwidth]{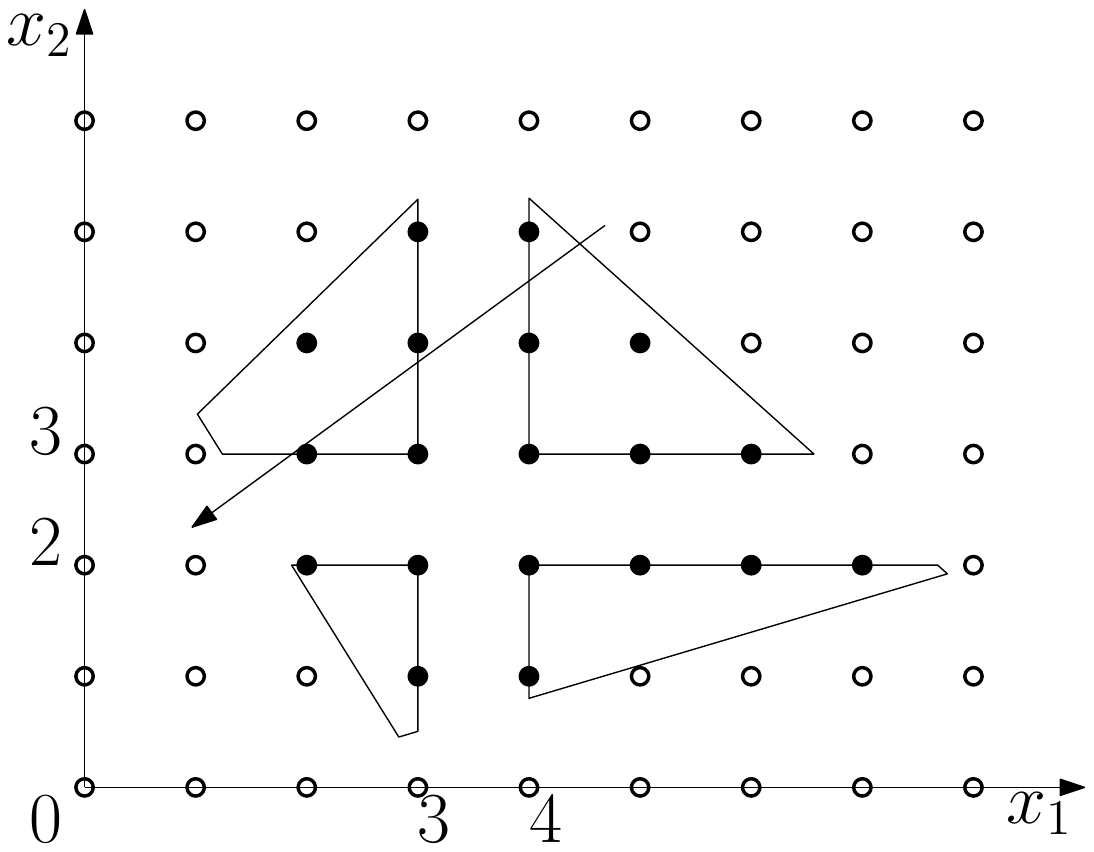}
  \caption{An example of fathom}
  \label{fig:5.3}
\end{figure}

\begin{figure}[!ht]
  \centering
  \includegraphics[width=0.5\textwidth]{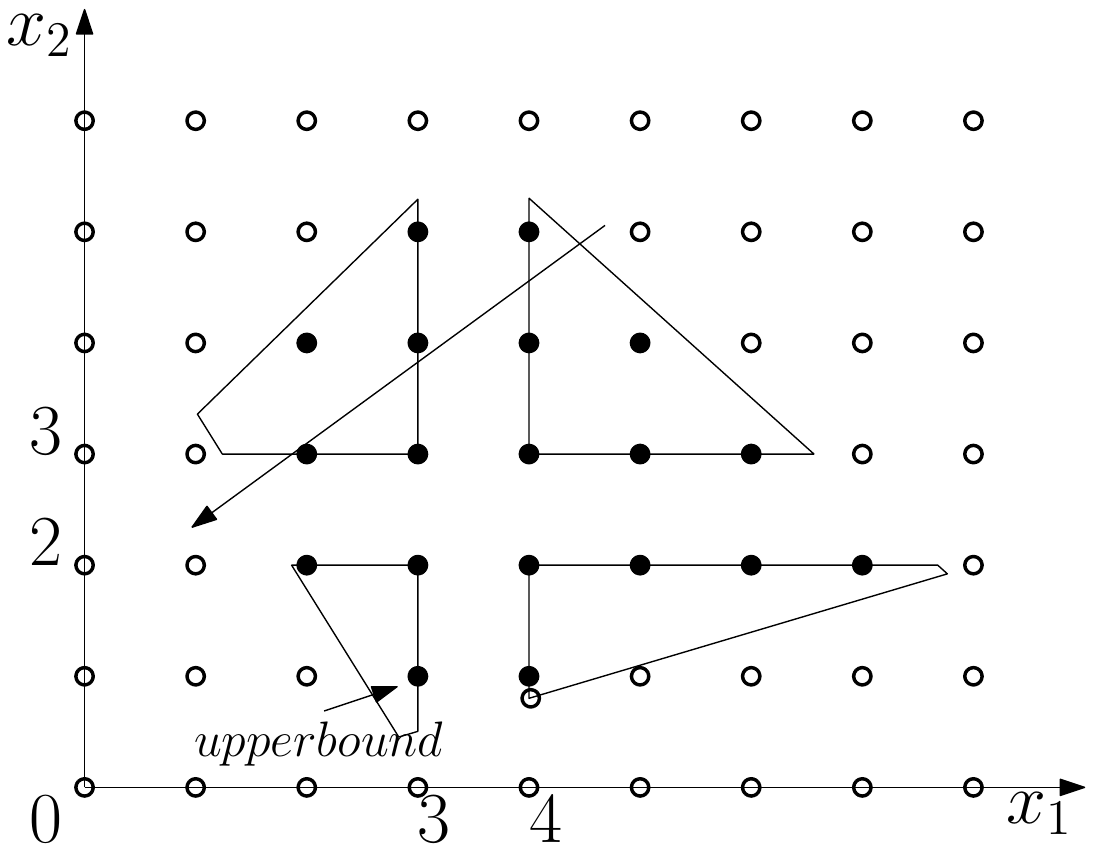}
  \caption{An example of high lower bound}
  \label{fig:5.4}
\end{figure}

In the pruning step, we can prune off a subproblem from the search tree under the following three cases.
\begin{description}
\item[Case 1] If the solution of the relaxation satisfies all the constraints of the candidate problem, then this solution is a feasible solution of the candidate problem; thus, this solution is the optimum solution of the candidate problem. If this happens, we say that this candidate problem is \emph{fathomed}. The lower bound and its solution will then be used to update the upper bound and the incumbent, and this candidate problem will be pruned off. This is the base case of the divide and conquer strategy.
\item[Case 2] If the lower bound is bigger than the current upper bound, that candidate problem will be pruned off, because any objective value of this candidate problem will be bigger than the upper bound.
\item[Case 3] If a candidate problem has no feasible solution, it will also be pruned off, since any further restrictions (via branching) will also be infeasible.
\end{description}
Case~3 is trivial; Case~1 and Case~2 are illustrated in Figure~\ref{fig:5.3} and Figure~\ref{fig:5.4}. 
In Figure~\ref{fig:5.3}, the arrow is the direction of the optimization. The subproblem at the upper right corner will be fathomed after solving an LP relaxation.
In Figure~\ref{fig:5.4}, suppose the current upper bound is defined by a feasible solution of the subproblem at the lower left corner (as labeled). After solving an LP relaxation for the subproblem at the lower right corner, we will get a non-integral solution. The objective value of the solution will be greater than the current upper bound.

If a subproblem can not be pruned off, we need to branch on that subproblem.

\begin{figure}[!ht]
  \centering
  \includegraphics[width=0.65\textwidth]{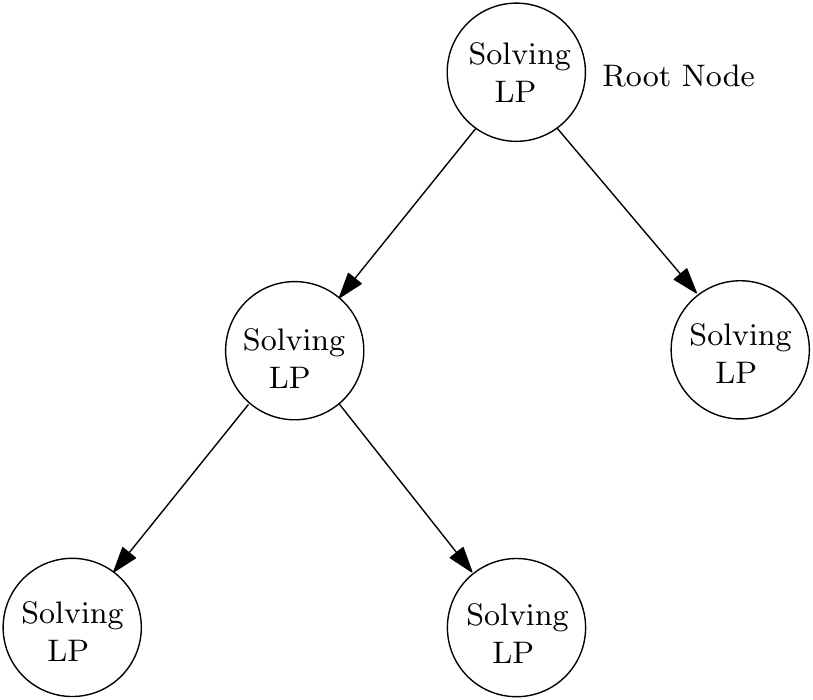}
  \caption{An example of search tree}
  \label{fig:5.5}
\end{figure}

A branch and bound algorithm will keep iterating these three steps, and terminate when no candidate problems are available. Figure~\ref{fig:5.5} is a typical search tree of a branch and bound algorithm. After all the candidate problems are considered, the last incumbent is the optimum solution of the original problem. In this method, not all the feasible solutions are enumerated, but the complete space of the feasible solutions is searched and the exact optimum solution is found. Thus, the branch and bound method is not a heuristic method.

\subsection{Strategies in the Branch and Bound Method}
\label{sec:bac.1.3}
The framework of the branch and bound method is very flexible. It is just a general method and does not specify the details in any of the three steps. Thus different techniques can be applied to each step.

\subsubsection*{Branching techniques}
A branching method is given in the above section. If $x$ is a binary variable, like $s_{i}$ in \eqref{eq:3.3} which can only be assigned the value of $0$ or $1$, we can have two candidate problems: one with the additional constraint of $x = 0$, the other with $x = 1$. In both of these two techniques, a problem is divided into two subproblems (so-called \emph{dichotomic} branching). The search tree will be a binary tree, like Figure~\ref{fig:5.5}. A problem can also be divided into more subproblems (so-called \emph{polytomic} branching); the resulting search tree will be a multiway tree. For more details of the branching techniques, please refer to~\cite{Clau,Lee,Murty}.

\subsubsection*{Branching variable selection}
In the branching step, it is important to carefully select the branching variable. Several methods are introduced in~\cite{Lee}. Murty~\cite{Murty} suggests that if several such variables are available, we usually choose the one that will produce the highest lower bound. The reason is that this strategy can reduce the gap between the upper and lower bounds, and thus can increase the chance of finishing the algorithm earlier.

\subsubsection*{Bounding techniques}
In the bounding step, we usually solve a relaxation for the lower bound. There are also other types of relaxation, for instance, Lagrangian relaxation (see~\cite{Lemar} for details). A good upper bound at the beginning of the algorithm can help prune more subproblems. Different heuristic algorithms are available. It is also important to find a good heuristic algorithm for a specific problem. 

\subsubsection*{Candidate problem selection}
One strategy for choosing the candidate problem is to choose the one which has the least lower bound because this candidate problem has the greatest chance that its optimum objective value is smaller than any lower bound of the other candidate problems. This strategy is called the \emph{best first search} strategy. \emph{Breadth first search} strategy and \emph{depth first search} strategy are also introduced in~\cite{Clau}.

There are many strategies available in every step other than the above mentioned ones. The choices depend on the characteristics of a specific problem. The performance of an algorithm depends on having good lower and upper bounds. It is better that the lower and upper bounds are close to the optimum objective value. The tighter the bounds are, the more we can prune off; but computing tighter bounds usually means more computational effort.

\subsection{An Algorithm Prototype}
\label{sec:bac.1.4}
The branch and bound method is only an algorithm skeleton, and has to be filled out for each specific problem. An algorithm prototype can be stated as follows.
\begin{enumerate}
\item[(1)] \emph{Initialization.} Solve a relaxation of the original problem to compute the lower bound of the optimum objective value. If the solution of the lower bound satisfies all the constraints of the original problem, the optimum solution of the original problem is found, and the algorithm terminates. If there is no feasible solution for the relaxed problem, there is no feasible solution for the original problem. If neither of these cases happens, find a feasible solution for the original problem with a heuristic algorithm and set the upper bound and the incumbent, or just set the upper bound to positive infinity. Finally, initialize an empty tree, and let the original problem be the root.
\item[(2)] \emph{Problem Selection.} If the tree is empty, the algorithm terminates. If there is an incumbent, it is the optimum solution of the original problem. If not, the original problem is infeasible. If the tree is not empty, select and remove a candidate problem from the tree with a selection rule.
\item[(3)] \emph{Branching.} Divide the selected candidate problem into a set of new candidate problems with a branching rule. The new candidate problems are the children of the original problem in the search tree.
\item[(4)] \emph{Bounding and Pruning.} For each new candidate problem generated in step~$(3)$, compute the lower bound. If the candidate problem satisfies any pruning criteria in Section~\ref{sec:bac.1.2}, discard it. If this candidate problem is fathomed, then update the upper bound and the incumbent with the value and the solution of this lower bound. After the upper bound is updated, discard any candidate problem with a lower bound which is greater than the current upper bound. If this candidate problem is not fathomed, put it into the tree. After processing all the new candidate problems, go to step~$(2)$.
\end{enumerate}

The order of these steps can vary, and the strategies in every step can be different from the ones in this prototype. Two typical branch and bound algorithms, eager and lazy branch and bound, are described in~\cite{Clau}.

\subsection{More about Branch and Bound}
\label{sec:bac.1.5}
To apply the branch and bound method, one needs to develop a specific algorithm for a specific problem. An algorithm can have good performance on one problem, but it can have poor performance on another.  For a large scale discrete and combinatorial optimization problem, finding a good feasible solution for the upper bound at the beginning of the algorithm is a ``key issue''~\cite{Clau}. The branch and bound method does not reduce the theoretical time complexity of the original problem. In the worst case, the search tree contains every feasible solution as a leaf. For a large scale problem, the computation load is usually too heavy for a single processor computer. Thus, parallel computers are usually employed for large scale problems, and, for this purpose, many parallel branch and bound algorithms have been developed.

\section{The Cutting Plane Method}
\label{sec:bac.2}
\begin{figure}[!ht]
  \centering
  \includegraphics[width=0.5\textwidth]{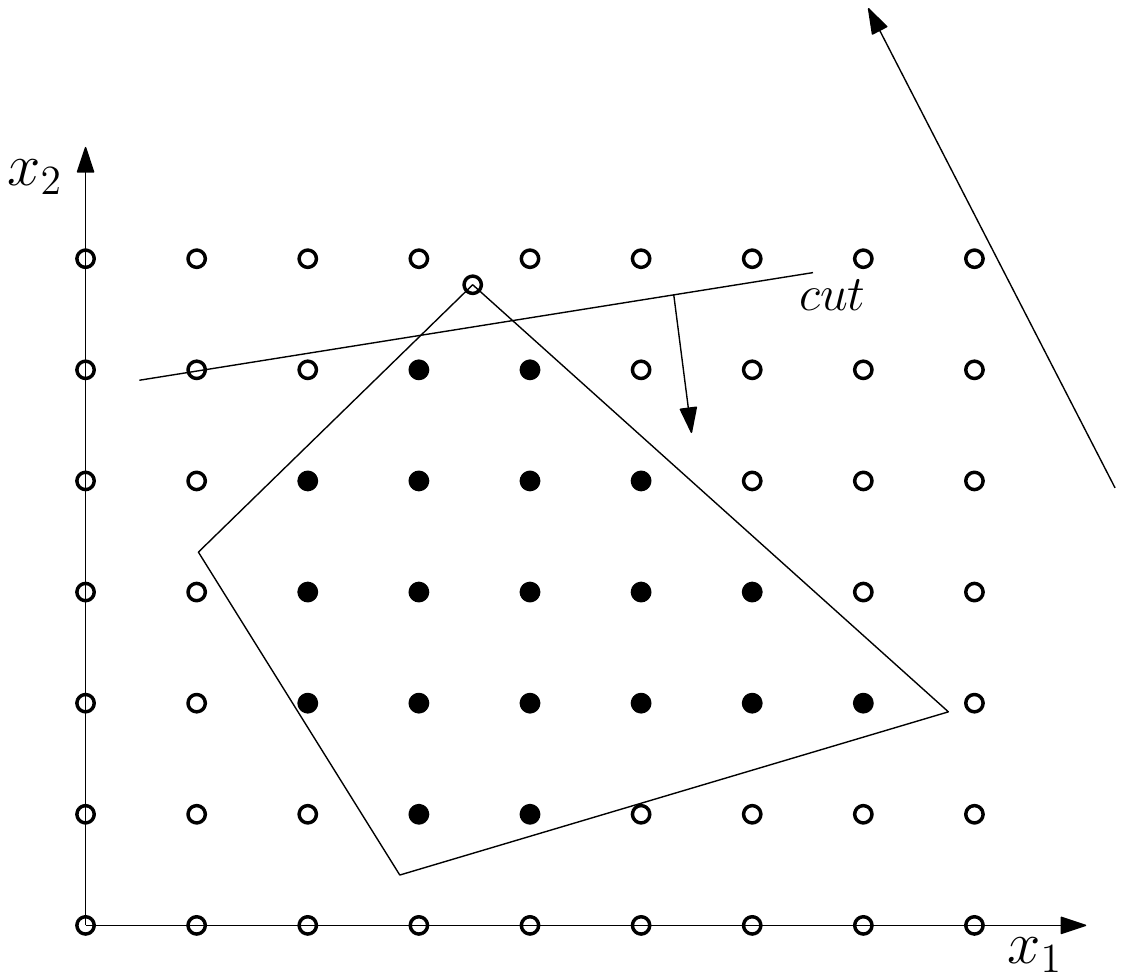}
  \caption{An example of a cutting plane}
  \label{fig:5.6}
\end{figure}

\begin{figure}[!ht]
  \centering
  \includegraphics[width=0.5\textwidth]{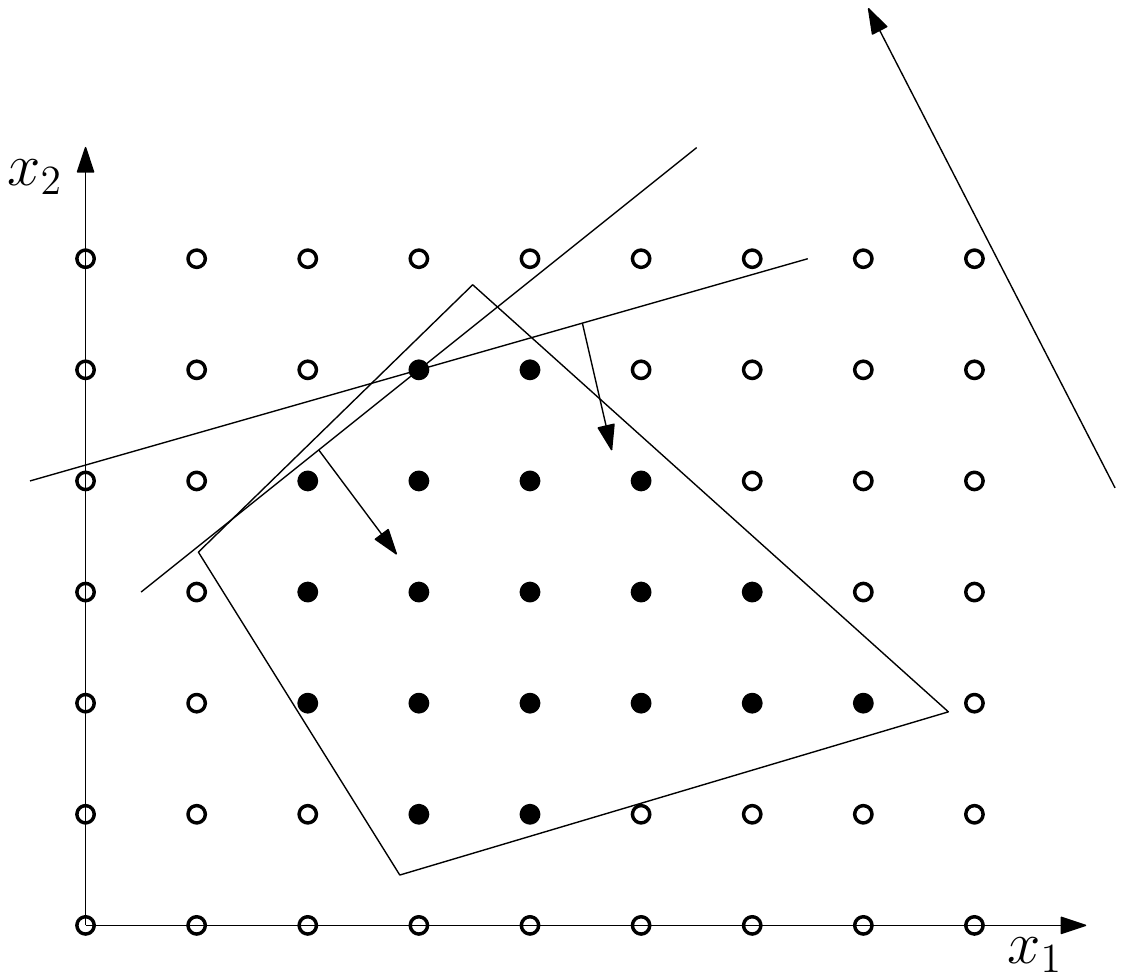}
  \caption{An example of cutting planes intersecting on the optimal solution}
  \label{fig:5.7}
\end{figure}

The \emph{cutting plane} method is an approach to improve the non-integral solution of the LP relaxation of an integer linear programming problem. After solving an LP relaxation, the optimal solution may not be integral. The cutting planes (or cuts) are the constraints which are satisfied by the original integer programming problem, but violated by the non-integral solution. We can find some such cutting planes and add them to the original problem to reduce the feasible region of the LP relaxation; ideally, the solution of the new LP relaxation will be closer to the integral optimal solution (as shown in Figure~\ref{fig:5.6}). The cutting plane algorithms will keep repeating the process of adding cuts and solving the linear relaxation until the integral optimal solution is found. As shown in Figure~\ref{fig:5.7}, if some cutting planes intersect on the integral optimal solution of the integer program and define an optimal vertex, the integer program will be solved.

In fact, it is hard to find the optimum integral solution by the cutting plane method itself, although it has solved some problems successfully. Several types of cutting planes for general integer programs have been proposed, for example, Chv\'atal-Gomory cuts, knapsack cuts, and lift-and-project cuts. For more details about the cutting plane method, please refer to~\cite{JEMcut,Nemh}.

These general cutting planes may not work well for some problems. Some problem-specific cuts can be developed for a specific problem. Recall that the halfspace depth problem is a hitting set problem. The optimal solution of \eqref{eq:3.3} is a hitting set of all IISs in \eqref{eq:3.2}. For an IIS $C$, we can formulate an IIS inequality \eqref{eq:mip.iisinq}. We can use such constraints as cuts for \eqref{eq:3.3}; this reduces the problem of finding cuts to that of finding IISs. As mentioned in Section~\ref{sec:mfs.2}, an IIS corresponds to a vertex of the infeasible system's alternative polyhedron. \cite{Khach} shows that generating all IISs of an infeasible system is NP-hard.

\section{The Branch and Cut Method}
\label{sec:bac.3}
In a branch and bound algorithm, we can apply the cutting plane method to every node in the search tree. Then, the branch and bound approach can be sped up dramatically. The combination of the branch and bound and the cutting plane method is called the \emph{branch and cut} method. For more details about branch and cut, please refer to~\cite{Hillier,JEMbc1,JEMbc2,Junger}.

\subsection{Parallel Branch and Cut}
\label{sec:bac.3.1}
As mentioned above, a branch and bound algorithm may not be fast enough for a large problem. The same is unfortunately true for branch and cut algorithms. Therefore, parallel branch and cut algorithms have been developed. A natural idea of parallelizing the branch and cut method is assign each subproblem to a processor. The drawback of this idea is that the workload for different processors could be significantly different, because some subproblems could be solved with little effort and some could be hard to solve, so we would not have good efficiency. A better idea is the master-slave paradigm. The master processor maintains the search tree, and delivers subproblems to the slave processors when possible. This idea has better efficiency than the first one, although it has more inter-processor communications and the master processor may be overly busy. For more information about parallel branch and cut methods, please refer to~\cite{Clausen1,Ralphs3,Ralphs1,Ralphs2}.

In this chapter we introduced the branch and bound and cutting plane method for the general mixed integer programs. In the next chapter we will introduce our branch and cut algorithm for the data depth problem.

\chapter{The Branch and Cut Algorithm}
\label{chap:alg}

\section{The Algorithm}
\label{sec:alg.1}
We develop a branch and cut algorithm for the halfspace depth problem. In this algorithm, we first use Chinneck's heuristic algorithm to find a feasible solution and set up the upper bound with this solution. We then initialize the search tree. The main part is iteratively selecting and processing a problem from the tree. After all the problems in the tree are processed, the optimum solution of the problem is found. In every iteration, we repeatedly solve an LP relaxation and add hitting set cuts. If the subproblem can be solved, it will be pruned off, otherwise, it will be divided into two subproblems.

The top level algorithm is shown in Algorithm~\ref{alg:6.1}, and Algorithm~\ref{alg:6.2} and Algorithm~\ref{alg:6.3} are subroutines of this algorithm.

\renewcommand{\algorithmicrequire}{\textbf{Input:}}
\renewcommand{\algorithmicensure}{\textbf{Output:}}
\renewcommand{\algorithmiccomment}[1]{/$\ast$ #1 $\ast$/}

\begin{algorithm}
  \caption{${\ensuremath{\mbox{\sc HalfSpaceDepth}}}(S,p)$}
  \label{alg:6.1}
  {\setlength{\baselineskip}{1.5\baselineskip}
    \begin{algorithmic}[1]
      \REQUIRE A set $S$ of points and a point $p$ in $\mathbb{R}^{d}$.
      \ENSURE The halfspace depth of $p$.
      \STATE  Generate an infeasible system $S_{1}$ and the corresponding integer program $S_{2}$ with the input.
      \STATE  Find an MIN IIS COVER $c$ of $S_{1}$ with Chinneck's heuristic algorithm.
      \IF     {$c == 0 \quad or \quad c == 1$}
      \STATE  \textbf{return} $c$
      \ENDIF
      \STATE  $upperbound = c$
      \STATE  Initialize the search tree with  $S_{2}$ as the root.
      \WHILE  {the tree is not empty}
      \STATE  Remove a problem $P$ from the search tree. \COMMENT{with depth first search strategy}
      \STATE  Call ${\ensuremath{\mbox{\sc BoundandCut}}}(P)$.
      \IF     {$P$ is infeasible $or$ the objective value $\geq upperbound$}
      \STATE  \textbf{continue}
      \ELSIF  {the subproblem is fathomed}
      \STATE  $upperbound = $ the objective value of $P$
      \STATE  \textbf{continue}
      \ELSE   [the subproblem is not fathomed]
      \STATE  Call ${\ensuremath{\mbox{\sc Branch}}}(P)$ and add the new subproblems into the search tree.
      \ENDIF
      \ENDWHILE
      \STATE \textbf{return} $upperbound$
    \end{algorithmic}
  }
\end{algorithm}

\begin{algorithm}
  \caption{${\ensuremath{\mbox{\sc BoundandCut}}}(P)$}
  \label{alg:6.2}
  {\setlength{\baselineskip}{1.6\baselineskip}
    \begin{algorithmic}[1]
      \REQUIRE An integer program $P$.
      \ENSURE A solution of $P$. \COMMENT{may not be integral}
      \STATE  Solve a linear program relaxation of $P$.
      \IF     {$P$ is infeasible $or$ the objective value $\geq upperbound$}
      \STATE  Report the result and \textbf{return}
      \ELSIF  {the solution is not integral}
      \REPEAT
      \STATE  Generate some hitting set cuts (the details are explained in Section~\ref{sec:alg.2}), add them into $P$, and resolve $P$.
      \UNTIL  {the solution is not sufficiently improved $or$ the solution is integral $or$ no cuts can be generated}
      \ENDIF
      \STATE  \textbf{return} the solution
    \end{algorithmic}
  }
\end{algorithm}

\begin{algorithm}
  \caption{${\ensuremath{\mbox{\sc Branch}}}(P)$}
  \label{alg:6.3}
  {\setlength{\baselineskip}{1.6\baselineskip}
    \begin{algorithmic}[1]
      \REQUIRE An integer program $P$.
      \ENSURE Subproblems of $P$
      \STATE  Identify the set of constraints $S_{1}^{n}$ in $S_{1}$ that correspond to the constraints in $P$.
      \STATE  Solve an elastic program of $S_{1}^{n}$. Find the constraint that has the best chance to be in the MIN IIS COVER of $S_{1}$ using the observations in Section~\ref{sec:heur.2} (the details of branching variable selecting are explained in Section~\ref{sec:alg.2}).
      \STATE  Let $s_{b}$ be the binary variable that corresponds to that constraint. Divide $P$ into two new subproblems by fixing $s_{b}$.
      \STATE  \textbf{return} the two new subproblems.
    \end{algorithmic}
  }
\end{algorithm}

\section{Special Techniques in This Algorithm}
\label{sec:alg.2}

\subsubsection*{Initial Heuristic Algorithm}
Chinneck's heuristic algorithm is very fast and accurate. Most of the time this heuristic finds an optimum solution. Hence, we will have a very good upper bound at the beginning of the branch and cut algorithm. The heuristic in~\cite{Chinneck} is used, because it is faster according to Chinneck.

\subsubsection*{IIS Hitting Set Cuts}
We apply IIS hitting set cuts for the problem. They are problem-specific cutting planes. 

\subsubsection*{Pseudo-Knapsack Technique for Generating Cuts}
In order to generate cuts that are violated by the current solution of the LP relaxation, we use a pseudo-knapsack technique to find as many binary variables as possible with a summation smaller than $1$ (Note that the binary variables will become continuous variables in the LP relaxation, and with bounds $0 \leq x_{i} \leq 1$ for any variable $x_{i}$). After solving a LP relaxation, the binary variables are ranked according their values in increasing order. Select the first $k$ variables ($k$ is maximal) such that the summation of them is smaller than $1$. Find the IISs in the corresponding constraints of these variables. Such an IIS must give a violated cutting plane for the current solution of the LP relaxation.

In fact, identifying the maximum set of binary variables is not a true knapsack problem, because in this problem the cost and the value of an item (a binary variable) are the same. The greedy method in the above paragraph will give the optimal solution of this pseudo-knapsack problem. We can prove this by contradiction. Suppose $\{a_{1}, a_{2}, \ldots, a_{n}\}$ is the set of the values of the binary variables in increasing order, the greedy method identifies the first $k$ items, and a better algorithm identifies a set $J$ of $j$ items ($j > k$). The sum of any $k + 1$ items in $J$ is greater or equal to $\sum_{i=1}^{k+1}a_{i}$, because if $J$ contains any items that are different from the items in $\{a_{1}, a_{2}, \ldots, a_{k + 1}\}$, any of those different items would be greater or equal to $a_{k + 1}$. Hence, the sum of the items in $J$ would be greater than $1$, noting that $\sum_{i=1}^{k+1}a_{i} > 1$. Therefore, a better algorithm cannot exist.

This technique is used in one of the two hitting set cut generators we implemented.

\subsubsection*{Branching Variable Selecting Rule}
When selecting the branching variable, we mimic the technique in Chinneck's heuristic algorithm. After solving an elastic program of $S_{1}^{n}$, we estimate the drop of SINF that each constraint can give by Observation~6 in Section~\ref{sec:heur.2}. The constraint $b$ which can give the most significant drop has the best chance to be a member of the MIN IIS COVER of $S_{1}$ according to Observation~3 in~\cite{Chinneck}. The binary variable $s_{b}$ that corresponds to $b$ is selected as the branching variable.

\subsubsection*{Candidate Problem Selection}
By fixing $b$, we get two new candidate problems, one with $s_{b} = 1$, the other with $s_{b} = 0$. In the problem selection step, a depth first strategy is used and the problem with $s_{b} = 1$ is selected as the new problem to process. As mention in Section~\ref{sec:mip.2}, fixing the binary variable $s_{b}$ to $1$ has the effect of removing constraint $b$ from $S_{1}$. As the algorithm continues to dive in the problem tree, $S_{1}$ will usually become feasible quickly due to the accuracy of Chinneck's algorithm. At that point, the candidate problem will be fathomed because the optimum objective value will be $0$, an integral solution. This strategy will hopefully keep the depth of the search tree small, so then we would have a good chance to have small search tree for the whole problem.

\section{A Binary Search Idea}
\label{sec:alg.bin}

As we mentioned in Section~\ref{sec:mip.2}, we can not check the accuracy of the solutions of the MIPs. However, we can find an accurate solution for a problem by solving several MIPs. The idea is as follows:

\subsection{New MIP Formulation}
\label{sec:alg.bin.mip}
In this idea, the MIP~\eqref{eq:3.3} needs to be changed to the following form:
\begin{eqnarray}
  \label{eq:alg.bin}
  \textrm{minimize} \qquad -\epsilon & & \nonumber \\
  \textrm{subject to} \qquad \qquad \qquad \qquad
  \sum_{j = 1}^{n} s_{j} & \leq & guess \nonumber \\
  \sum_{i = 1}^{d} (A_{1}^{i} - A_{p}^{i}) x_{i} + s_{1}M & \geq & \epsilon \nonumber \\
  \sum_{i = 1}^{d} (A_{2}^{i} - A_{p}^{i}) x_{i} + s_{2}M & \geq & \epsilon \\
  \vdots  & & \vdots \nonumber \\
  \sum_{i = 1}^{d} (A_{n}^{i} - A_{p}^{i}) x_{i} + s_{n}M & \geq & \epsilon \nonumber \\
  s_{j} & \in & \{0 , 1\} \qquad \forall j \in \{1, 2, \ldots, n\} \nonumber \\
  \epsilon & \geq & 0 \nonumber \\
  - \infty \leq & x_{i} & \leq + \infty \qquad \forall i \in \{1, 2, \ldots, d\} \nonumber
\end{eqnarray}
In this formulation $\epsilon$ is a variable, and there is also one more constraint in which $guess$ is a value we want to test the depth against. If the optimal value of the objective function is $0$, $guess$ is smaller than the depth of point~$A_{p}$.

\subsection{The Binary Search Algorithm}
\label{sec:alg.bin.alg}
In this algorithm we need to modify our branch and cut algorithm, Algorithm~\ref{alg:6.1}, before using it as a subroutine. The subroutines of formulating MIPs and Chinneck's heuristic are separated from the original Algorithm~\ref{alg:6.1}. It will only solve a MIP, and it will terminate as soon as it finds a feasible solution which gives a nonzero $\epsilon$, because a nonzero $\epsilon$ implies that $guess$ is no less than the depth of $A_{p}$. The binary search algorithm, shown in Algorithm~\ref{alg:binmip}, maintains a cut pool containing the cutting planes generated in the early Algorithm~\ref{alg:6.1} subroutines. The cuts will be used as indexed cuts for later Algorithm~\ref{alg:6.1} subroutines.

\begin{algorithm}
  \caption{${\ensuremath{\mbox{\sc HalfSpaceDepthWithBinarySearch}}}(S,p)$}
  \label{alg:binmip}
  {\setlength{\baselineskip}{1.5\baselineskip}
    \begin{algorithmic}[1]
      \REQUIRE A set $S$ of points and a point $p$ in $\mathbb{R}^{d}$.
      \ENSURE The halfspace depth of $p$.
      \STATE  Generate an infeasible system $S_{1}$ with the input.
      \STATE  Find an MIN IIS COVER $c$ of $S_{1}$ with Chinneck's heuristic algorithm.
      \IF     {$c == 0 \quad or \quad c == 1$}
      \STATE  \textbf{return} $c$
      \ENDIF
      \STATE  Initialize a cut pool.
      \STATE  $upperbound = c$; $lowerbound = 1$
      \STATE  $guess = \lfloor (upperbound + lowerbound)/2 \rfloor$
      \WHILE  {$lowerbound < upperbound$}
      \STATE  Formulate an MIP $S_{2}$ with $guess$
      \STATE  Call ${\ensuremath{\mbox{\sc HalfSpaceDepth}}}(S_{2})$ and add the newly generated cuts into the cut pool.
      \IF     {$\epsilon == 0$ $or$ the MIP is infeasible }
      \STATE  $lowerbound = guess + 1$
      \STATE  $guess = \lfloor (upperbound + lowerbound)/2 \rfloor$
      \ELSE
      \STATE  $upperbound = guess$
      \STATE  $guess = \lfloor (upperbound + lowerbound)/2 \rfloor$
      \ENDIF
      \ENDWHILE
      \STATE \textbf{return} $upperbound$
    \end{algorithmic}
  }
\end{algorithm}

In this chapter we introduced our branch and cut algorithm for the half\-space depth problem. Due to the problem in the MIP formulation, we can not guarantee an accurate solution with Algorithm~\ref{alg:6.1}. Therefore we developed a binary search idea to find the accurate solution. In the next chapter we will introduce the implementation details of our algorithm.

\chapter{Implementation}
\label{chap:impl}
Our branch and cut algorithm is implemented with the BCP library from the \textbf{COIN-OR} project~\cite{Coin}, along with the Osi, Clp and Cgl libraries from this project. In this chapter we will introduce these libraries and how our branch and cut algorithm is implemented with them. For the binary search algorithm, we just make some adjustments of the branch and cut algorithm, and use it as a subroutine.

\section{BCP}
\label{sec:impl.1}
BCP is an open source Branch-Cut-Price framework. \emph{Pricing} is another technique for solving integer programs~\cite{Barhart,Savelsb}. Our algorithm is a branch and cut algorithm, thus we only use the branch and cut part of BCP. BCP is a set of C++ classes and functions which manage the search tree. It does not contain any LP solver or cutting plane generator. The Osi (Open Solver Interface) library is used as the interface between BCP and an LP solver. Clp (COIN-OR linear programming) is used as the LP solver in our implementation. Some commercial LP solvers, like CPLEX or Xpress, might be faster than Clp, but we want other researchers to have easy access to our codes. Of course, it is possible to change the code in order to use other LP solvers thanks to Osi. Cgl (Cut Generation Library) is a collection of cut generators, which is used to generate cutting planes for BCP.

BCP only handles minimization problems, since a maximization can be easily transferred into a minimization problem. BCP is designed for parallel execution in the master slave paradigm; it also supports sequential execution. One philosophy of BCP is black box design: the users do not need to know the implementation details. To use BCP, in principle we only need to know its interfaces and the parameters. If one wants to use some techniques which BCP does not provide, one can ``open the box'' and edit the source code and recompile the library; this could be an obstacle because BCP currently lacks good documentation.

\subsection{Structure of BCP}
\label{sec:impl.1.1}
BCP has four independent computational modules: Tree Manager, Linear Programming (LP), Cut Generator, and Variable Generator.
\begin{description}
\item [The Tree Manager Module] This module is the master process. It initializes the problem and manages the search tree. It will keep track of all processes and distribute subproblems to the slave processes.
\item [The Linear Programming (LP) Module] This module is the most important part of BCP. It is a slave module, and the tree manager module can manage several LP modules during the parallel execution. The LP module does more than just solving a LP relaxation, but also applies cutting plane method, selects branching variables, and branches the subproblems. It performs all the branch, bound, and cut jobs. 
\item [The Cut Generator Module] This module will generate cutting planes for the LP module based on an LP solution.
\item [The Variable Generator Module] This module will generate variables for the LP module during the pricing process.
\end{description}

Only the first two modules are used in our implementation. The cut generator module is necessary when the work load or the required memory for generating cutting planes is big. In our algorithm the cuts can be generated in a short time. Hence, it is better to generate the cuts in the LP module. As we mentioned before, different strategies can be applied to one operation in the branch and bound method. We need to specify a set of parameters for BCP to use some specific techniques that have been implemented in BCP.

\subsection{Parallelization}
\label{sec:impl.1.2}
The design of the independent modules make parallelization easy. The modules communicate by passing messages. BCP supports both MPI and PVM protocol. In our implementation MPI is used. The tree manager process maintains a list of subproblems, and assigns a subproblem to an LP process when it is idle.  A single list of candidate problems is maintained by the tree manager process. This is a bottleneck for parallelization. Because of the limitations of memory and CPU power of a single node, the tree manager process can only manage a limited number of LP and other processes.

For more details about BCP, please refer to the BCP manual~\cite{Bcpman}, although some contents are out of date.

\section{Implementation Details}
\label{sec:impl.2}
The codes of this algorithm are based on the example \textbf{BAC}~\cite{Margot} written by Margot. We also implemented Chinneck's heuristic algorithm~\cite{Chinneck} with Osi and Clp, and two cut generators to generate the hitting set cuts.

\subsection{The MPS File Generator}
\label{sec:impl.dtl.mps}
A C++ class template to generate an MPS (Mathematical Programming System, a text file format for linear programs) file is implemented. This generator will read the input data from a text file, then generate an MPS file of the MIP and an MPS file of the infeasible system. Some ANOVA (Analysis of Variance) applications need duplicated points in the input. In this case, the same duplicated data are either all inside the halfspace or all outside the halfspace when finding the depth of a point. Therefore, the binary variables associated with these data will be one or zero simultaneously. When formulating the MIP, we keep only one of the duplicated constraints and assign a weight of the number of the duplicated constraints to the associated binary variable in the objective function.

\subsection{The Cut Generators}
\label{sec:impl.2.1}
The cut generator will receive the solutions of an LP relaxation from the LP process and generate cutting planes based on these solutions. We implemented two hitting set cuts generators with Cgl. One is based on the idea in~\cite{David}, the other is based on the idea in the appendix of~\cite{Bremner1}

Bremner, Fukuda, and Rosta developed a primal-dual algorithm for the halfspace depth problem in~\cite{David}. Their algorithm is to find the minimum traversal of all MDSs in the input data set. They developed a library to generate the MDSs (recall that MDSs are the same as IISs), and that library is based on Avis' \textbf{Lrslib} library~\cite{Avis}. We use this MDS generating library to generate IISs for our algorithm in the first cut generator.

In this cut generator, we use the pseudo-knapsack technique in Section~\ref{sec:alg.2} to find a set of binary variables. Then we identify the set of points in the input data set that correspond to the binary variables. This set of points are then used as the input for the MDS generate library to generate a set of IISs. Finally we formulate a set of cutting planes in the form of \eqref{eq:mip.iisinq}, one for each IIS found.

The paper \cite{Bremner1} gives an idea to generate a \emph{basic infeasible subsystem (BIS)} of an infeasible system. Given an infeasible system $Ax \geq b$ where $A \in \mathbb{R}^{n \times d}$ and $b \in \mathbb{R}^{n}$, the basic infeasible subsystem is an infeasible subsystem of cardinality no more than $d + 1$. To find a basic infeasible subsystem, the idea is to apply phase~1 of the two phase simplex method~\cite{Chvatal} by solving the following LP:
\begin{eqnarray}
  \label{eq:impl.phase1}
  \textrm{minimize} \qquad  x_{0} \quad \quad & & \nonumber \\
  \textrm{subject to} \qquad
  Ax +x_{0} & \geq & b
\end{eqnarray}
After getting the optimal solution, the set of tight constraints corresponds to a basic infeasible subsystem of $Ax \geq b$. For more details about the basic infeasible subsystem, please refer to~\cite{Bremner1}. The basic infeasible subsystem may not be irreducible if it contains a degenerated IIS whose cardinality would be smaller than $d + 1$, nevertheless it defines a cutting plane.

At every node of search tree in our algorithm, we have a unique infeasible system by removing the inequalities that are associated with the binary variables which have been fixed to one. Hopefully we can identify a different basic infeasible subsystem for each node.

BCP does not support global cuts currently. Any cuts added to a subproblem are only available to its children. This is unfortunate for us, since all hitting set cuts are globally valid. On the other hand, keeping too many cuts can slow down each node (Bremner, Fukuda, and Rosta~\cite{David} observed adding all IIS cuts made solving the LP relaxation very slow).

\subsection{The Tree Manager Process}
\label{sec:impl.2.2}
The tree manager process is the central process. It initializes the algorithm and manages the search tree. After the algorithm terminates, it will report the final results. The tree manager process performs the following functions:
\begin{itemize}
  \item Read the integer problem and the infeasible system from MPS files.
  \item Compute an initial upper bound for the integer program with the heuristic algorithm applied on the infeasible system.
  \item Initialize the integer problem.
  \item Initialize the search tree.
  \item Send the problem to the LP process(es).
  \item Receive solutions and update the best one.
  \item Receive the data and cuts that subproblems will need in the future.
  \item Receive requests from LP process(es) and send a subproblem.
  \item Receive branching information, and branch on the processed subproblem.
  \item Keep tracking the upper bound, and inform the LP process(es) when it is updated.
  \item Print the final results.
\end{itemize}

\begin{figure}[!ht]
  \centering
  \includegraphics[width=0.7\textwidth]{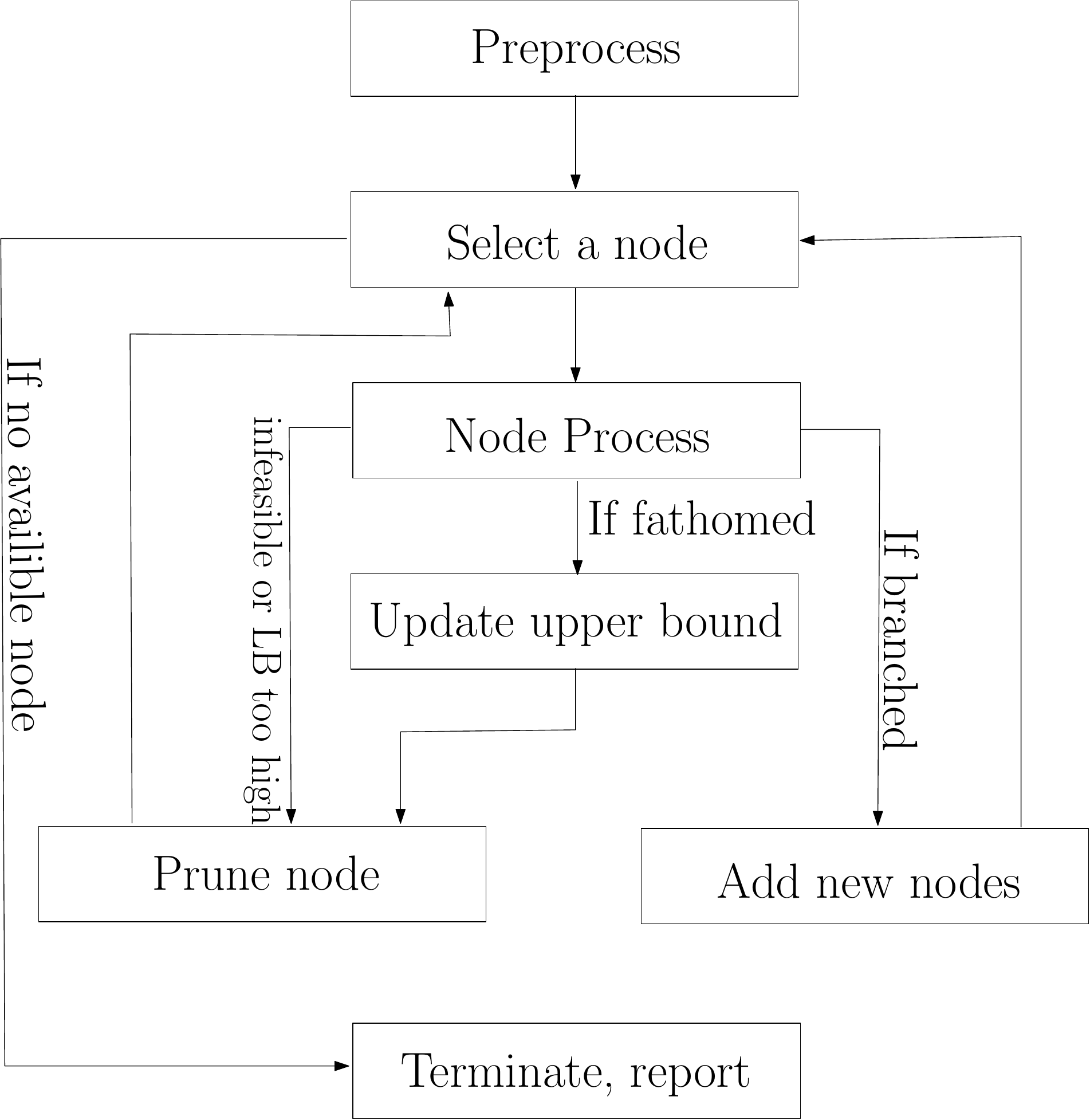}
  \caption{The Tree Manager Process}
  \label{fig:7.1}
\end{figure}

The work flow of the tree manager process is shown in Figure~\ref{fig:7.1}.

\subsection{The Linear Programming Process}
\label{sec:impl.2.3}
The LP process will receive a subproblem from the tree manager process and perform the branch and cut work on the subproblem. This process performs the following functions:
\begin{itemize}
  \item Initialization.
  \item Receive the problem from the tree manager process.
  \item Set up LP solver.
  \item Request new subproblem.
  \item Receive a subproblem and related data.
  \item Solve an LP relaxation.
  \item Test feasibility and fathoming.
  \item Generate cutting planes with cut generators and add the cuts into the subproblem.
  \item When necessary, choose branching object and apply branching strategy.
  \item Send results, related data, and cuts to tree manager process. 
\end{itemize}

\begin{figure}[!ht]
  \centering
  \includegraphics[width=0.7\textwidth]{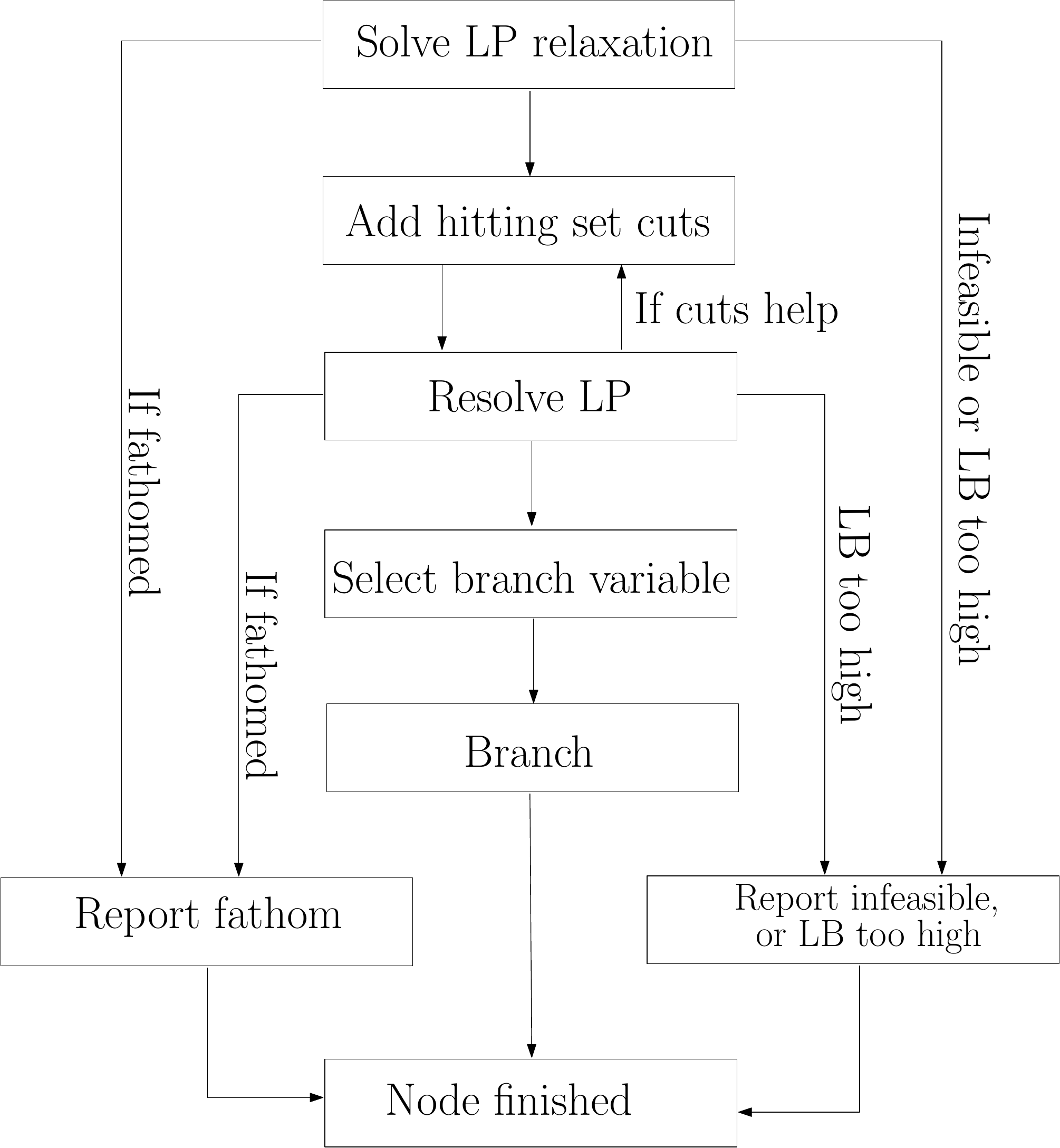}
  \caption{The Linear Programming Process}
  \label{fig:7.2}
\end{figure}

The work flow of the linear programming process is shown in Figure~\ref{fig:7.2}.

In this process, if the cut generator based on the MDS generator is used, cutting planes could always be found based on a solution of the LP relaxation. In each iteration we test how much the cutting planes improve the solution. If the objective value of the solution is improved by some amount, new cuts will be generated and another iteration will be done. The cut generator based on the phase~1 simplex method is easy to use. At each node one cut is generated. Other cut generators in Cgl, like the Gomory cut generator, are also used.

The cutting plane generated in each node are sent back to the TM process, and will be sent to the current node's children.

We also apply the simple rounding heuristic in~\cite{Margot} to the solution of the LP relaxation. This heuristic rounds the value of the integral variables to integers. In fact this heuristic does not work well most of the time. We only apply it when current search tree level is greater than $7$ and current iteration is greater than $5$.

\subsection{Parameters}
\label{sec:impl.2.4}
The parameters can be passed to the algorithm by a text file. We can specify the maximum running time of the algorithm. We can also specify the numerical precision. In this algorithm we set the granularity and integer tolerance to $10^{-12}$. We can also choose the branching strategy. In this algorithm, we need to disable the default strong branching strategy in order to use the greedy branching rule. To use the candidate problem selection strategy in Section~\ref{sec:alg.2}, we set the tree search strategy  to the depth first search and the child preference to dive down. Many other parameters can also be specified. The BCP documentation~\cite{Bcpdoc} contains the full list of parameters.

In this chapter we introduced the BCP library and the details of our implementation. In the next chapter we will present experimental results.

\chapter{Computational Experiments}
\label{chap:test}
Our algorithms have been tested on a Myrinet/4-way cluster that consists of dual socket SunFire x4100 nodes which are populated with 2.6 GHz dual-core Opteron 285 SE processors and 4 GB RAM per core. We set the CPU time limit to 60 minutes in these tests. For readability, we relegate most of the raw experimental results to an appendix, and report only a summary in this chapter.

\section{Numerical Issues}
\label{sec:test.numissu}

In practice, if the value of the $\epsilon$ in the MIP is too small compared with the coefficients of the constraints, the linear programming solver would round it to zero. Our solution is scaling the data items, and making the norms similar and relatively small. For a few data sets, the depth values reported by our algorithm with different strategies or parameters are different (with a difference of $1$). This could be caused by bugs in our codes or bugs in BCP, but we also suspect this is due to some numerical issue.

\section{Results for Random Generated Data}
\label{sec:test.rand}
The data sets tested in this section are a subset of the data sets used in~\cite{David}, and they are randomly generated. For every data set we compute the depth of the first point, which is the origin. For all the tests in this section, the $\epsilon$ of the MIP~\eqref{eq:3.3} is set to $0.00001$. Comparing with the results of the primal-dual algorithm and the binary search algorithm, the depth values computed with our branch and cut algorithm (with BIS cut generator) are accurate. Therefore the $\epsilon$ is small enough.

\subsection{Comparing Branching Rules and Tree Search Strategies}
\label{sec:test.rand.bran}

We first test our algorithm with the first hitting set cut generator in Section~\ref{sec:impl.2.1}, the one implemented with the MDS generating library, and with the greedy branching rule (see Section~\ref{sec:alg.2}). Table~\ref{tab:test.cutmds-d5}, Table~\ref{tab:test.cutmds-d10}, Table~\ref{tab:test.cutmds-n50}, and Table~\ref{tab:test.cutmds-sd5} give the performance on $4$ group of data sets. We generate $10$ cuts in one iteration of the LP process. If the objective value is improved by $0.001$, the LP process will do another iteration. When the MDS cut generator is used, most of the CPU time is spent on cutting plane generation. If more cuts are generated in one iteration, the algorithm will be slowed down, but will be more memory efficient.

\begin{figure}[!ht]
  \centering
  \includegraphics[width=0.9\textwidth]{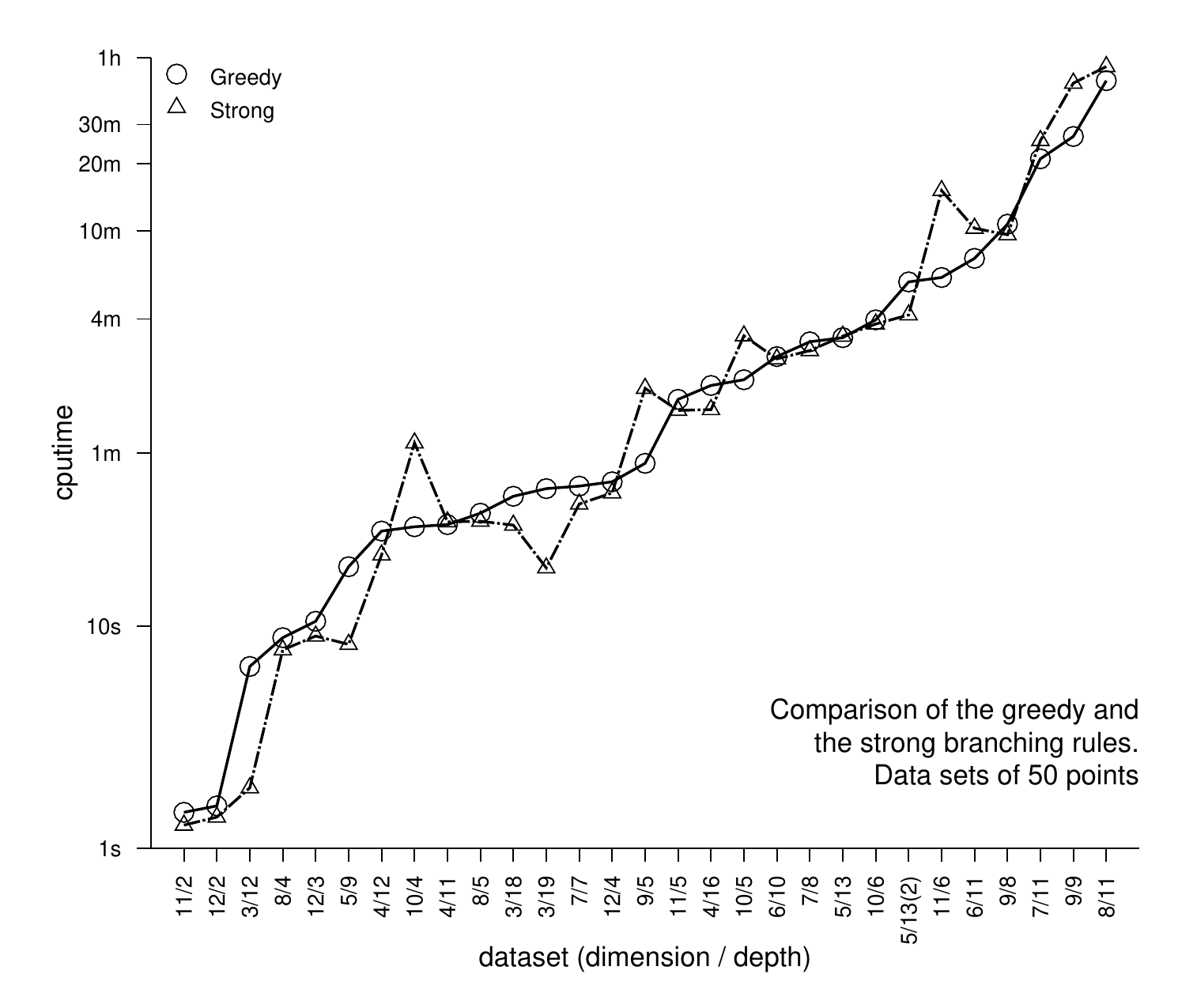}
  \caption{Comparison of different branching rules}
  \label{fig:test.bran-n50}
\end{figure}

\begin{figure}[!ht]
  \centering
  \includegraphics[width=0.9\textwidth]{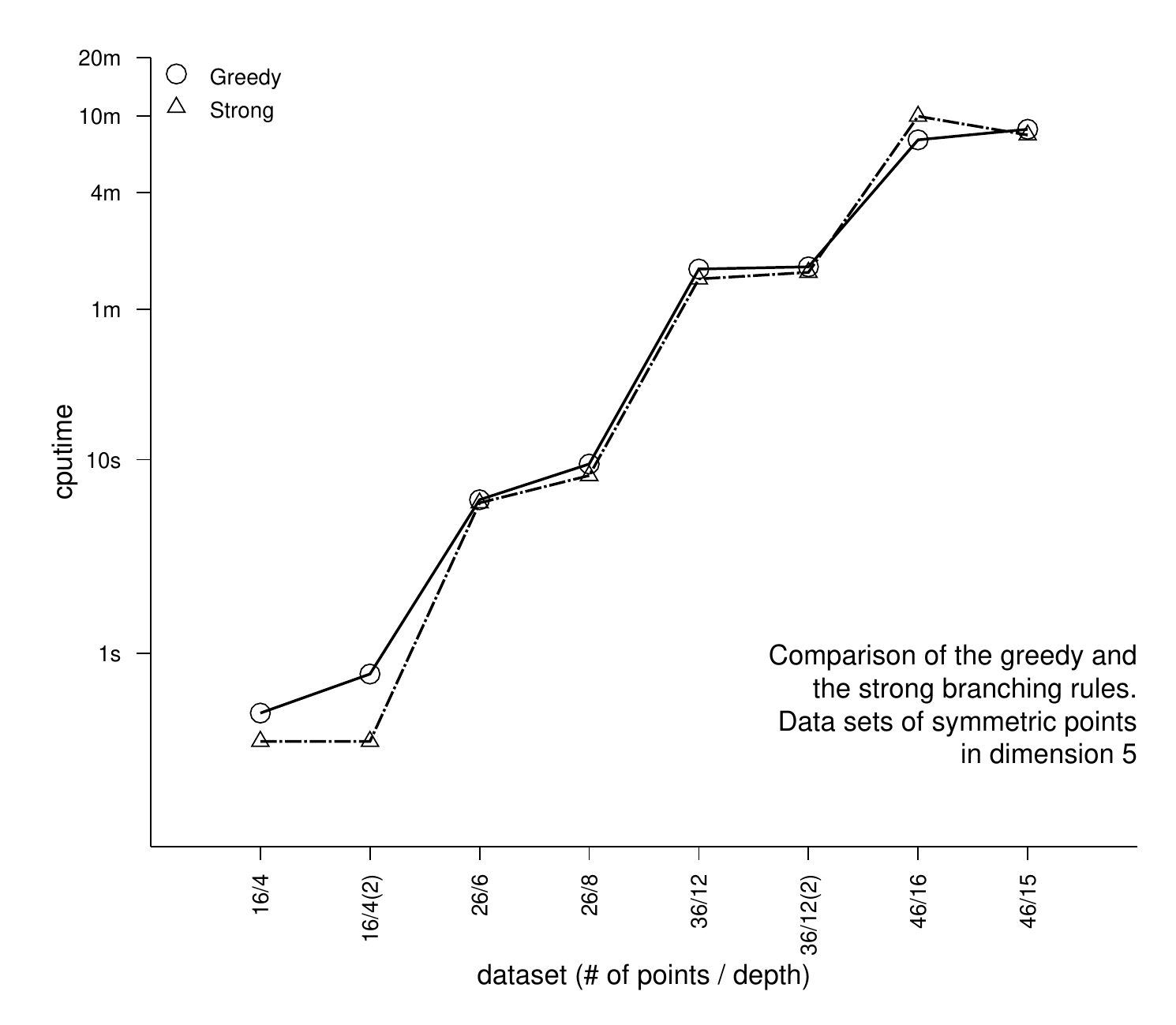}
  \caption{Comparison of different branching rules}
  \label{fig:test.bran-sd5}
\end{figure}

Table~\ref{tab:test.bran-d5}, Table~\ref{tab:test.bran-d10}, Table~\ref{tab:test.bran-n50}, and Table~\ref{tab:test.bran-sd5} give the performance when the default strong branching rule in BCP is used. In Figure~\ref{fig:test.bran-n50} and Figure~\ref{fig:test.bran-sd5} we compare the performance of the default strong with the greedy branching rules. Figure~\ref{fig:test.bran-n50} and Figure~\ref{fig:test.bran-sd5} show that strong branching gives better performance for most problems, probably because less search tree nodes are processed. For many difficult problems, the greedy branching works better. In those difficult cases, greedy branching spent much less time on branching, although more search tree nodes would be processed.

Table~\ref{tab:test.sel-d5}, Table~\ref{tab:test.sel-d10}, Table~\ref{tab:test.sel-n50}, and Table~\ref{tab:test.sel-sd5} give the performance of the best first candidate problem selecting rule. With this strategy, the performance is similar to that with depth first strategy. In these tests, strong branching and MDS cut generator are applied.

\subsection{Comparing Cut Generators}
\label{sec:test.rand.cut}

Table~\ref{tab:test.cutbis-d5}, Table~\ref{tab:test.cutbis-d10}, Table~\ref{tab:test.cutbis-n50}, and Table~\ref{tab:test.cutbis-sd5} give the performance of our algorithm compiled with the second hitting set cut generator in Section~\ref{sec:impl.2.1}, the one implemented with basic infeasible system idea. In these tests, the default strong branching in BCP is used. With the BIS cut generator, less CPU time will be used to generate cuts, and the algorithm has better overall performance, although the search tree is larger. The BIS cut generator uses floating point arithmetic, the same as the rest of the system. The MDS cut generator uses exact arithmetic which is required for the \textbf{Lrslib}. This is a factor which slows down the MDS cut generator.

In fact many cuts are generated repeatedly in the optimization process. The pseudo-knapsack idea in Section~\ref{sec:alg.2} can force the algorithm to generate a different cut each time, but the performance turns out to be worse, and more search tree nodes will be processed. With the pseudo-knapsack idea, if the algorithm generates a cut with a probability less than $1$ on each node, the performance will be improved to some extent, although still worse than that without pseudo-knapsack. We also observe that the pseudo-knapsack idea can make the algorithm faster when the greedy branching is applied. This suggests that the pseudo-knapsack idea interferes with the strong branching rule. The reason might be that this idea makes the values of the binary variables in the solution of LP relaxation closer to each other.

\begin{figure}[!ht]
  \centering
  \includegraphics[width=0.9\textwidth]{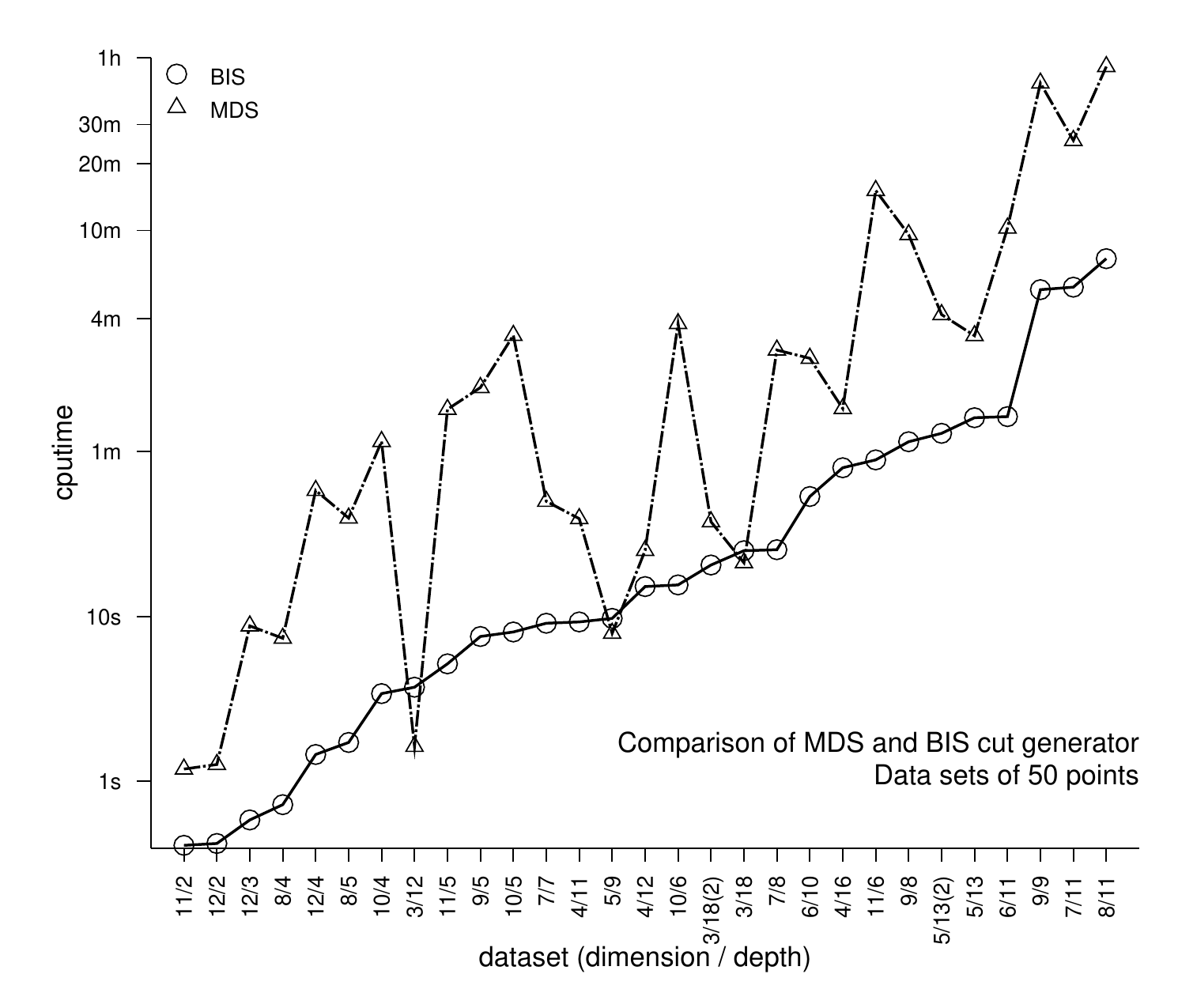}
  \caption{Comparison of different cutting plane generators}
  \label{fig:test.cut-n50}
\end{figure}

\begin{figure}[!ht]
  \centering
  \includegraphics[width=0.9\textwidth]{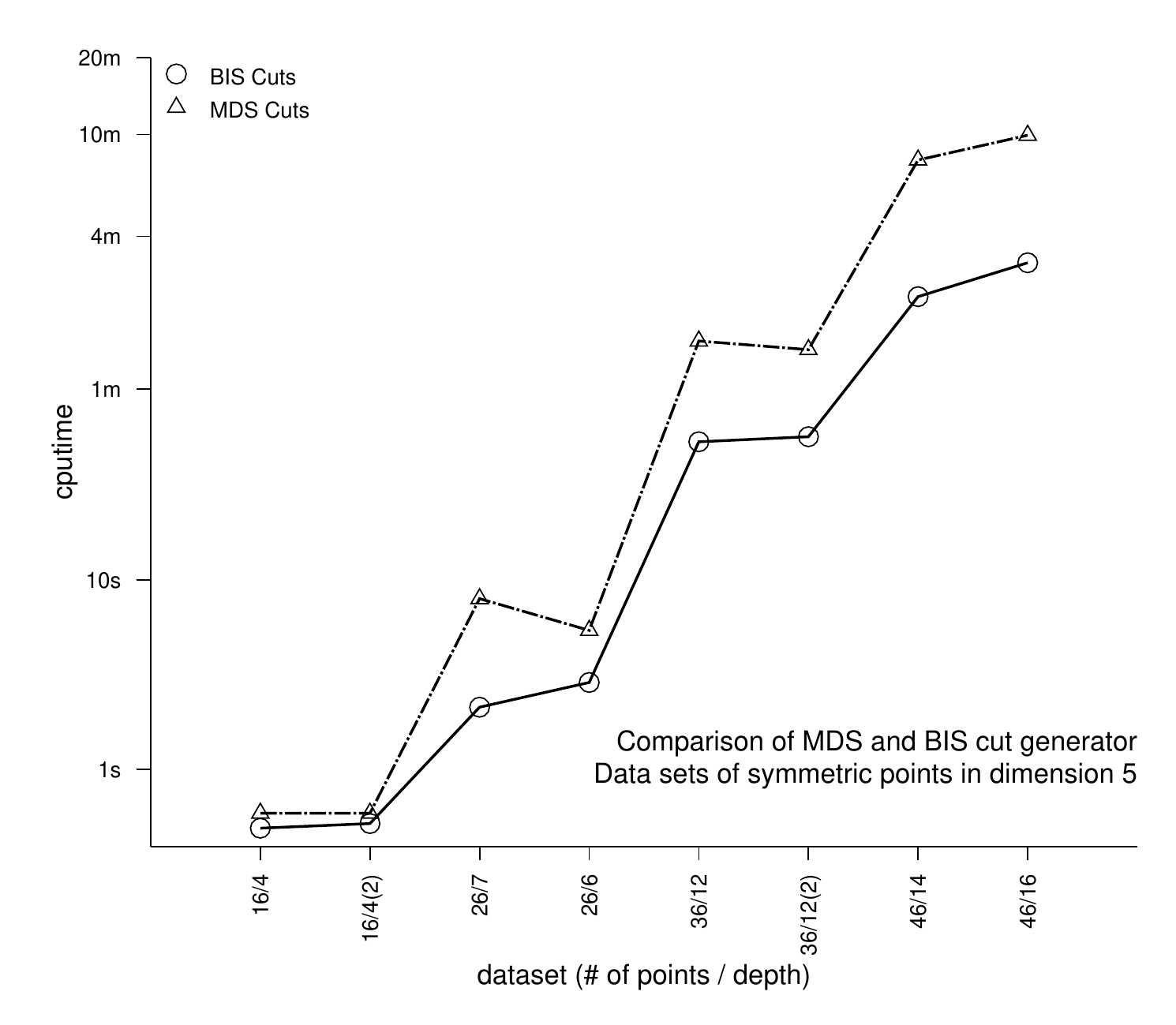}
  \caption{Comparison of different cutting plane generators}
  \label{fig:test.cut-sd5}
\end{figure}

In Figure~\ref{fig:test.cut-n50} and Figure~\ref{fig:test.cut-sd5} we compare the performance of our algorithm with the two different cutting plane generators. The general cut generators in Cgl can barely generate cuts for our algorithm, and do not improve the performance.

\subsection{Comparing Algorithms}
\label{sec:test.rand.alg}

\begin{figure}[!ht]
  \centering
  \includegraphics[width=0.9\textwidth]{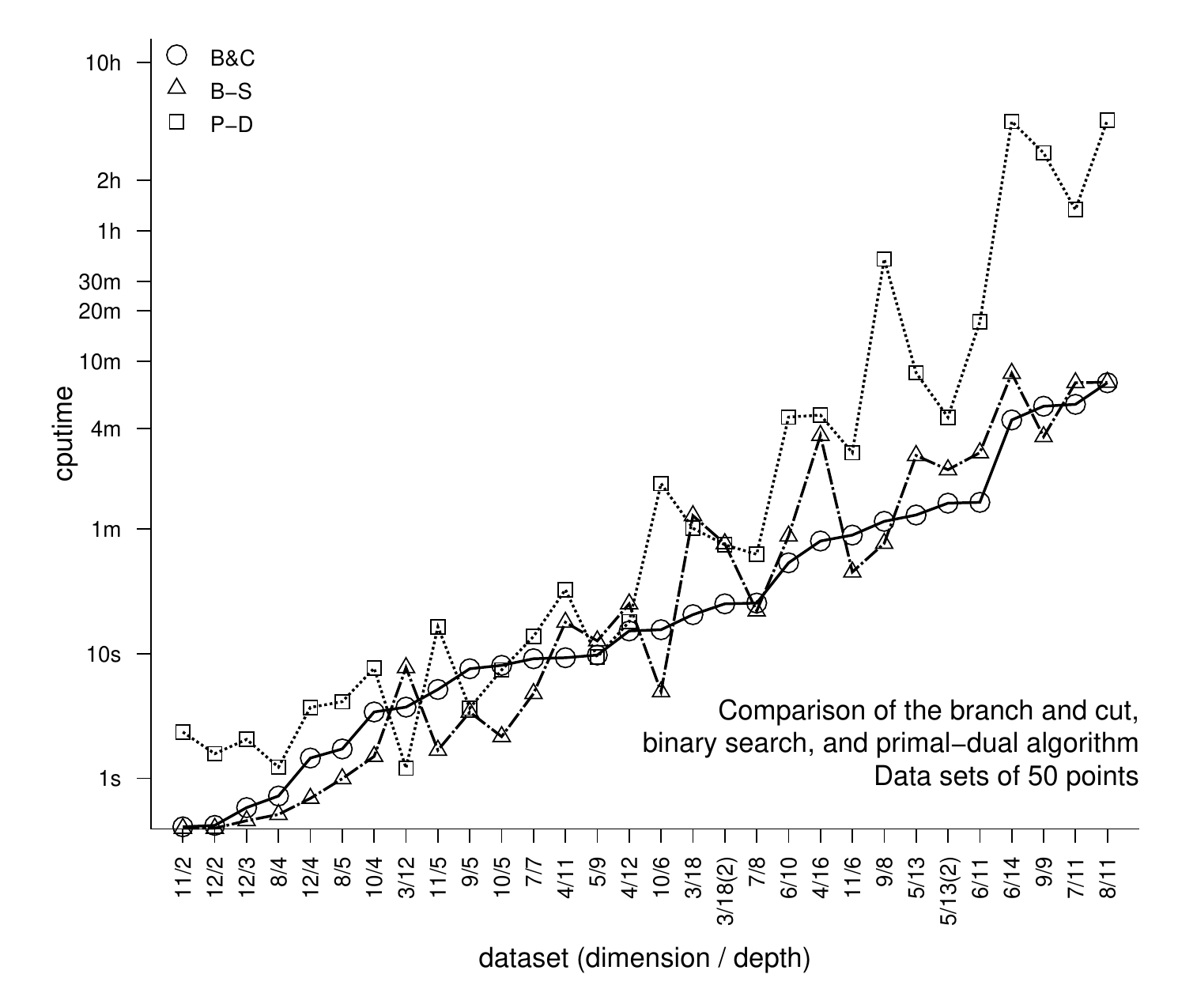}
  \caption{Comparison of different algorithms}
  \label{fig:test.bbp-n50}
\end{figure}

\begin{figure}[!ht]
  \centering
  \includegraphics[width=0.9\textwidth]{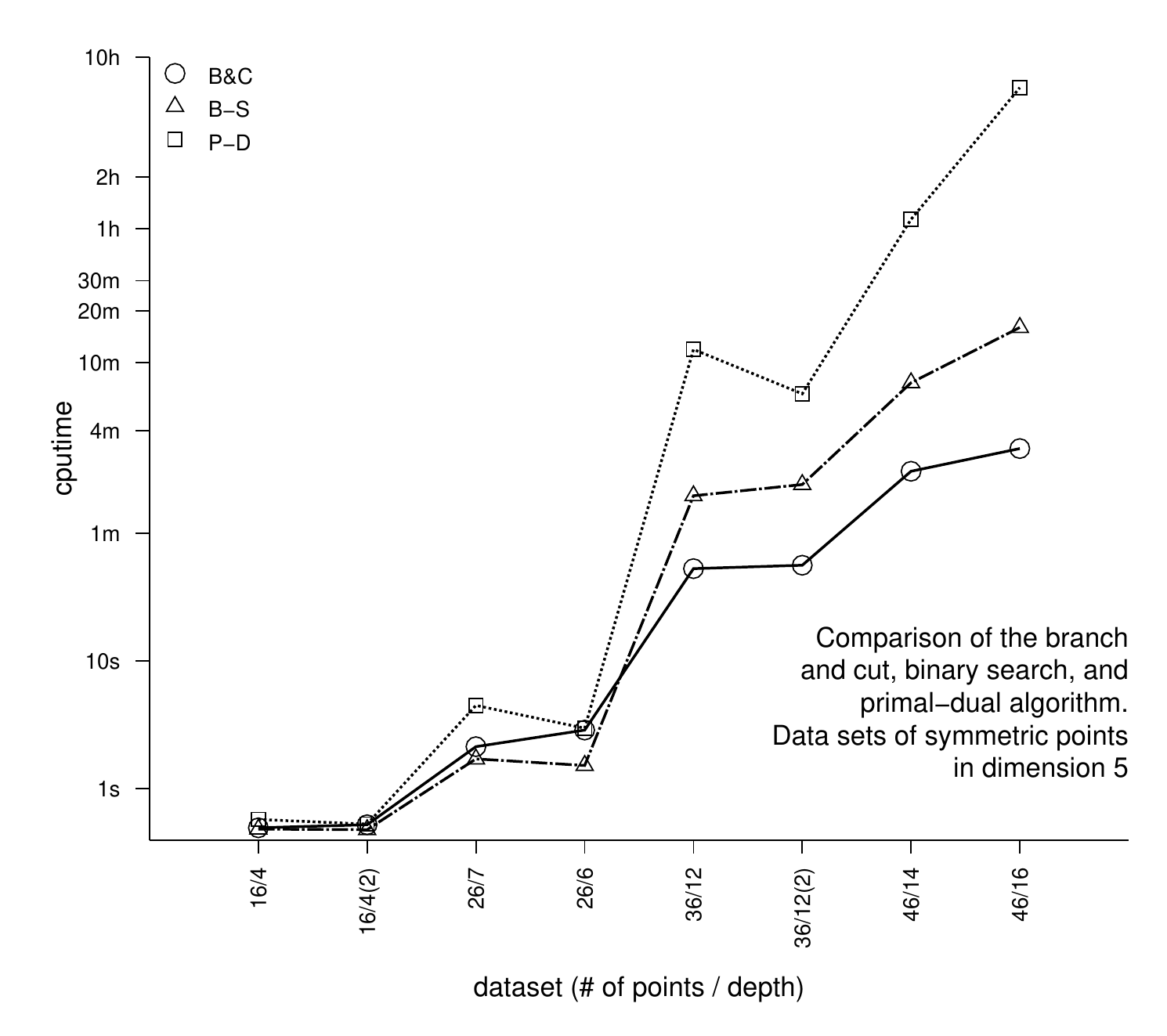}
  \caption{Comparison of different algorithms}
  \label{fig:test.bbp-sd5}
\end{figure}

The performance of the binary search algorithm in Section~\ref{sec:alg.bin.alg} is given in Table~\ref{tab:test.bin-d5}, Table~\ref{tab:test.bin-d10}, Table~\ref{tab:test.bin-n50}, and Table~\ref{tab:test.bin-sd5}. The time in the tables is the total time of solving all MIPs during the binary process. The performance of the primal-dual algorithm on the same data sets is given in Table~\ref{tab:test.pd-d5}, Table~\ref{tab:test.pd-d10}, Table~\ref{tab:test.pd-n50}, and Table~\ref{tab:test.pd-sd5}. The binary search algorithm does not perform too badly, but the primal-dual algorithm is very slow on some hard problems. In Figure~\ref{fig:test.bbp-n50} and Figure~\ref{fig:test.bbp-sd5} we compare the performance of the binary search algorithm, the primal-dual algorithm, and the branch and cut algorithm. The branch and cut algorithm works best most of the time. The performance of the binary search algorithm is actually quite fast (as well as being more numerically stable). Sometimes the binary search algorithm even works better than the branch and cut algorithm. The reason is that the MIPs for the binary search algorithm are usually easier to solve, and the tricks used in the binary search algorithm also help to speed up the algorithm. In contrast, the primal-dual algorithm can be slow on large problems.

\subsection{Parallel Execution}
\label{sec:test.rand.para}

\begin{figure}[!ht]
  \centering
  \includegraphics[width=0.9\textwidth]{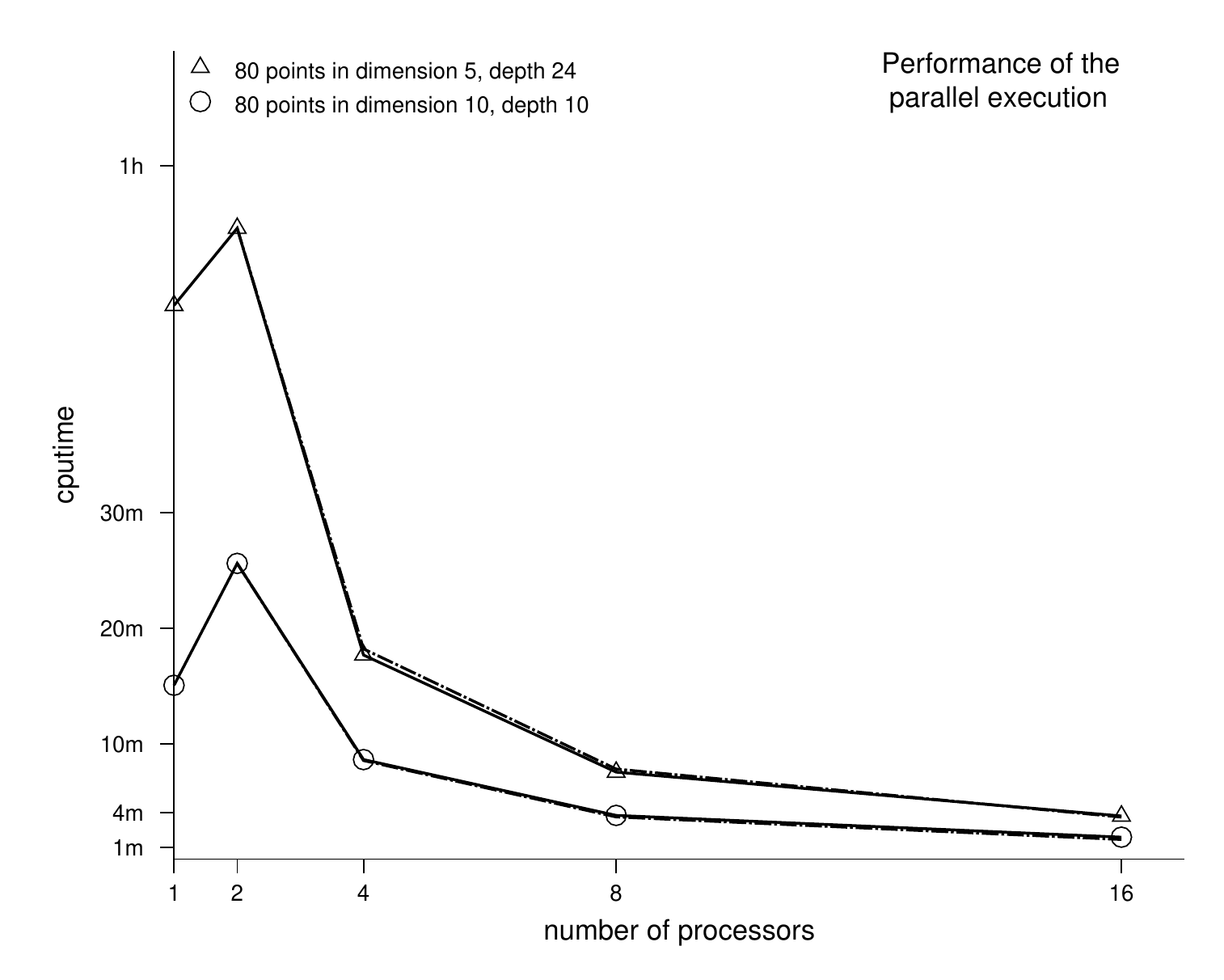}
  \caption{The Performance of Parallel Execution}
  \label{fig:test.paral}
\end{figure}

All the above tests are done with the sequential version of our algorithm. Some tests of parallel version of the branch and cut algorithm are given in Figure~\ref{fig:test.paral}. Two data sets are used to test the algorithm. The performance with one processor is the performance of the sequential version of the algorithm. When two processors are applied, one of them is used for the slave process (LP process), and when four processors are applied, three of them are used for the slave process. So we expect a speedup of $3$ for four processors, $7$ for eight processors, and so forth. The dashed line in the figure indicates the linear speedup with respect to number of LP processes. From the figure we can see that the speedup is almost linear.

\section{Results for ANOVA Data}
\label{sec:test.2}
The ANOVA data tested in this section are randomly generated according to a scheme in~\cite{Mizera}. The data points are in the form of $y_{i_{1} \ldots i_{m}}$. In this manuscript Mizera gives the following description:
\singlespace
\begin{quote}
  The index $i_{k}$ corresponds to a $k$-th \emph{factor}. The values $(1, 2, \ldots, I_{k})$ of this index are called \emph{factor levels} (apparently, [it] is only this $I_{k}$ which is technically of interest here) and correspond to the variants of the treatment that the factor represents. For instance, a factor may correspond to a kind of soil (say, there are three different ones involved in the experiment); another factor may represent the type of the fertilizer used (say, five different fertilizers used in the experiment); a datapoint then records the yield on the plot (or a plot; there may be more plots of this kind) corresponding to the particular combination of soil and fertilizer.
\end{quote}
\doublespace
There are some duplicated points in every data set. For every data set, the depth of the origin is computed. Table~\ref{tab:test.mip} gives a comparison of the performance of our algorithm with different integer program formulations, the simple MIP as~\eqref{eq:3.3} and the weighted MIP as described in Section~\ref{sec:impl.dtl.mps}. The sequential algorithm is used for these tests. The upper bound of the number of duplications in the data set is given in the third column. From the table we can see that the algorithm using the weighted MIP is much faster, because there are many fewer rows in the MIP.
\begin{table}[!htb]
  \centering
    \begin{tabular}[center]{|c|c|c|c|c|c|}
    \hline
    Point \# & Dimension & Duplication & Depth & Simple MIP & Weighted MIP \\
    \hline
    32 & 8 & 2 & 5 & 0.62 & 0.19 \\
    32 & 8 & 2 & 10 & 5.89 & 2.06 \\
    32 & 8 & 2 & 7 & 0.90 & 0.24 \\
    32 & 8 & 2 & 7 & 0.54 & 0.31 \\
    32 & 8 & 2 & 4 & 0.14 & 0.03 \\
    48 & 8 & 3 & 5 & 0.11 & 0.08 \\
    48 & 8 & 3 & 11 & 4.39 & 0.82 \\
    48 & 8 & 3 & 10 & 2.17 & 0.62 \\
    48 & 8 & 3 & 9 & 2.04 & 0.28 \\
    48 & 8 & 3 & 13 & 27.72 & 2.00 \\
    64 & 8 & 4 & 11 & 1.92 & 0.22 \\
    64 & 8 & 4 & 17 & 91.89 & 2.98 \\
    64 & 8 & 4 & 18 & 200.82 & 3.48 \\
    64 & 8 & 4 & 15 & 30.85 & 0.87 \\
    64 & 8 & 4 & 16 & 28.94 & 1.60 \\
    72 & 12 & 2 & 13 & 147.59 & 22.19 \\
    72 & 12 & 2 & 18 & 807.20 & 250.77 \\
    72 & 12 & 2 & 14 & 85.94 & 33.52 \\
    72 & 12 & 2 & 17 & 529.16 & 69.92 \\
    72 & 12 & 2 & 20 & outmem & 469.81 \\
    108 & 12 & 3 & 26 & outmem & 519.49 \\
    108 & 12 & 3 & 24 & outmem & 264.20 \\
    108 & 12 & 3 & 24 & outmem & 341.87 \\
    108 & 12 & 3 & 29 & outmem & 1435.35 \\
    108 & 12 & 3 & 22 & outmem & 105.49 \\
    144 & 12 & 4 & 33 & outmem & 1238.99 \\
    144 & 12 & 4 & 39 & outmem & 1760.49 \\
    144 & 12 & 4 & 40 & outmem & 1527.83 \\
    144 & 12 & 4 & 33 & outmem & 544.95 \\
    144 & 12 & 4 & 29 & outtime & 330.57 \\
    \hline
  \end{tabular}
  \caption{Performance with different integer program formulations}
  \label{tab:test.mip}
\end{table}

\chapter{Conclusions}
\label{chap:concl}

\section{Summary of the Work}
\label{sec:concl.sum}

We noted that the halfspace depth problem is equivalent to the maximum infeasible subsystem problem. We reviewed a heuristic algorithm suggested by Chinneck. We also reviewed the branch and cut paradigm for MIP problems. We developed a branch and cut algorithm for the halfspace depth problem. Based on this algorithm, we developed a second binary search algorithm for the halfspace depth problem. Two cut generators were also developed for our algorithm.

We evaluated different strategies for the branch and cut algorithm. We also compared the branch and cut algorithm with the binary search algorithm and the primal-dual algorithm, and concluded that the branch and cut algorithm is the fastest, although with some numerical issues. The binary search is slower, but still faster than the primal-dual algorithm and more stable. Fast cutting plane generators are important, because the BIS cut generator improves the performance dramatically. The strong branching rule is faster than the greedy branching rule on most of the tests, but the greedy branching rule is faster on many hard problems (i.e. those with large depth).

On ANOVA data sets, the duplicated constraints are removed with the weighted MIP formulation. With this modification, the algorithm solved all the problems we tested.

\section{Open Problems and Future Work}
\label{sec:concl.open}
In some applications, only the center of the data set is interesting. With the current algorithm we have to compute the depth of every data item in order to find the center. A fast algorithm for finding the center is open for future work. 

The idea for finding a proper $\epsilon$ described in Section~\ref{sec:mip.2} is not practical. Another open problem is a method to find a practical $\epsilon$ for MIP~\eqref{eq:3.3}. There may be an idea to solve an MIP based on the strict inequalities of system~\eqref{eq:3.1}. Then we do not need to consider $\epsilon$. The binary search algorithm does not require a value for $\epsilon$, and it can report a proper value for $\epsilon$. Ironically, this algorithm finds a proper value after solving the halfspace depth problem.

As we noticed in Section~\ref{sec:test.rand.cut}, the pseudo-knapsack idea slows down the strong branching when using the BIS cut generator. An idea for reducing redundant cut generation in the BIS cut generator that does not interfere with the strong branching would be interesting.

\singlespace
\bibliographystyle{amsplain}
\addcontentsline{toc}{chapter}{Bibliography}
\bibliography{thesis}
\appendix

\chapter{Testing Results}
\label{chap:apd}

\begin{table}[!htb]
  \centering
  \begin{tabular}[center]{|r|l|}
    \hline
    Num & The number of points in the input data set \\
    Dim & The dimension of the data items \\
    Suc & Success or not \\
    TD & The depth of the search tree \\
    UB & The upper bound found by the heuristic algorithm or the best\\
    &solution found by the algorithm if the algorithm does not\\
    &finish successfully \\
    Nod & The number of processed nodes in the search tree \\
    Tim & The CPU time used to find the solution \\
    Dep & The optimal objective value of the problem \\
    outtime & Running out of time\\
    outmem & Running out of memory\\
    $(*)$ & The depth value with this note are one larger than the correct\\
    &value\\
    \hline
  \end{tabular}
  \caption{Abbreviations used in this chapter}
  \label{tab:test.abb}
\end{table}

\begin{table}[!htb]
  \centering
  \begin{tabular}[center]{|r|l|}
    \hline
    d5 & A group of data sets in dimension 5\\
    d10 & A group of data sets in dimension 10\\
    n50 & A group of data sets, each set consists of 50 points\\
    sd5 & A group of data sets in dimension 5; the points are symmetric around\\
        & the origin in each set\\
    \hline
  \end{tabular}
  \caption{Data sets used for the tests}
  \label{tab:test.datasets}
\end{table}

\section{Results of the Branch and Bound algorithm}
\label{sec:apd.bac}

\subsection{Results of the Greedy Branching and MDS Cut Generator}
\label{sec:apd.bac.gre}

\begin{table}[!htb]
  \centering
  \begin{tabular}[center]{|c|c|c|c|c|c|c|c|}
    \hline
    Num & Dim & Suc & Tim & TD & Nod & UB & Dep \\
    \hline
    20 & 5 & yes & 0.36 & 9 & 183 & 4 & 4 \\
    20 & 5 & yes & 0.48 & 10 & 157 & 5 & 5 \\
    30 & 5 & yes & 2.19 & 12 & 553 & 6 & 6 \\
    30 & 5 & yes & 1.00 & 9 & 317 & 4 & 4 \\
    30 & 5 & yes & 5.49 & 13 & 1255 & 8 & 8$(*)$ \\
    40 & 5 & yes & 1.07 & 10 & 91 & 5 & 5 \\
    40 & 5 & yes & 26.77 & 16 & 3101 & 10 & 10 \\
    40 & 5 & yes & 48.20 & 16 & 5857 & 11 & 11$(*)$ \\
    50 & 5 & yes & 63.03 & 16 & 7505 & 10 & 10 \\
    50 & 5 & yes & 277.55 & 20 & 27061 & 14 & 14 \\
    50 & 5 & yes & 341.83 & 20 & 25825 & 15 & 15 \\
    60 & 5 & outmem &&&& 16 &\\
    60 & 5 & outmem &&&& 19 &\\
    60 & 5 & yes & 257.61 & 19 & 15191 & 13 & 13 \\
    70 & 5 & outmem &&&& 21 &\\
    70 & 5 & outmem &&&& 20 &\\
    70 & 5 & outmem &&&& 23 &\\
    80 & 5 & outtime &&&& 20 &\\
    80 & 5 & outmem &&&& 22 &\\
    80 & 5 & outmem &&&& 25 &\\
    90 & 5 & outmem &&&& 29 &\\
    90 & 5 & outmem &&&& 21 &\\
    90 & 5 & outmem &&&& 24 &\\
    \hline
  \end{tabular}
  \caption{Performance with the greedy branching rule, data set: d5}
  \label{tab:test.cutmds-d5}
\end{table}

\begin{table}[!htb]
  \centering
  \begin{tabular}[center]{|c|c|c|c|c|c|c|c|}
    \hline
    Num & Dim & Suc & Tim & TD & Nod & UB & Dep \\
    \hline
    50 & 10 & yes & 78.80 & 16 & 5431 & 5 & 5 \\
    50 & 10 & yes & 6.74 & 13 & 377 & 3 & 3 \\
    50 & 10 & yes & 604.58 & 19 & 45741 & 7 & 7 \\
    60 & 10 & yes & 96.16 & 16 & 5315 & 5 & 5 \\
    60 & 10 & yes & 325.15 & 17 & 15939 & 6 & 6 \\
    60 & 10 & yes & 32.29 & 15 & 1639 & 5 & 4 \\
    70 & 10 & yes & 478.59 & 17 & 16019 & 6 & 6 \\
    70 & 10 & outtime &&&& 10 &\\
    70 & 10 & yes & 1457.65 & 18 & 43105 & 7 & 7 \\
    80 & 10 & outtime &&&& 10 &\\
    80 & 10 & outtime &&&& 8 &\\
    80 & 10 & outtime &&&& 14 &\\
    90 & 10 & outtime &&&& 18 &\\
    90 & 10 & outtime &&&& 16 &\\
    90 & 10 & outtime &&&& 13 &\\
    110 & 10 & outtime &&&& 19 &\\
    120 & 10 & outimet &&&& 21 &\\
    120 & 10 & outtime &&&& 25 &\\
    130 & 10 & outtime &&&& 28 &\\
    \hline
  \end{tabular}
  \caption{Performance with the greedy branching rule, data set: d10}
  \label{tab:test.cutmds-d10}
\end{table}

\begin{table}[!htb]
  \centering
  \begin{tabular}[center]{|c|c|c|c|c|c|c|c|}
    \hline
    Num & Dim & Suc & Tim & TD & Nod & UB & Dep \\
    \hline
    50 & 3 & yes & 41.45 & 21 & 2949 & 19 & 19$(*)$ \\
    50 & 3 & yes & 6.57 & 13 & 373 & 12 & 12 \\
    50 & 3 & yes & 38.29 & 19 & 2743 & 18 & 18 \\
    50 & 4 & yes & 120.65 & 19 & 7203 & 16 & 16 \\
    50 & 4 & yes & 26.70 & 16 & 1719 & 12 & 12 \\
    50 & 4 & yes & 28.55 & 15 & 2025 & 11 & 11 \\
    50 & 5 & yes & 18.46 & 12 & 853 & 9 & 9 \\
    50 & 5 & yes & 197.99 & 20 & 20725 & 14 & 13 \\
    50 & 5 & yes & 352.73 & 20 & 41845 & 14 & 13 \\
    50 & 6 & outmem &  &  &  & 14 &  \\
    50 & 6 & yes & 449.99 & 19 & 58455 & 12 & 11 \\
    50 & 6 & yes & 162.71 & 17 & 14229 & 10 & 10 \\
    50 & 7 & yes & 1261.95 & 22 & 155239 & 13 & 11 \\
    50 & 7 & yes & 42.53 & 15 & 3169 & 7 & 7 \\
    50 & 7 & yes & 189.65 & 17 & 20883 & 8 & 8 \\
    50 & 8 & yes & 2835.12 & 21 & 290617 & 12 & 11 \\
    50 & 8 & yes & 8.85 & 13 & 891 & 4 & 4 \\
    50 & 8 & yes & 32.14 & 14 & 2677 & 5 & 5 \\
    50 & 9 & yes & 53.85 & 15 & 3871 & 6 & 5 \\
    50 & 9 & yes & 1590.48 & 20 & 149065 & 10 & 9 \\
    50 & 9 & yes & 641.22 & 19 & 66863 & 8 & 8 \\
    50 & 10 & yes & 27.92 & 15 & 1983 & 5 & 4 \\
    50 & 10 & yes & 128.03 & 17 & 9551 & 6 & 5 \\
    50 & 10 & yes & 237.98 & 18 & 16823 & 6 & 6 \\
    50 & 11 & yes & 369.11 & 19 & 24865 & 7 & 6 \\
    50 & 11 & yes & 104.49 & 18 & 7681 & 5 & 5 \\
    50 & 11 & yes & 1.45 & 13 & 83 & 2 & 2 \\
    50 & 12 & yes & 1.55 & 14 & 87 & 2 & 2 \\
    50 & 12 & yes & 10.49 & 15 & 581 & 3 & 3 \\
    50 & 12 & yes & 44.48 & 17 & 2705 & 4 & 4 \\
    \hline
  \end{tabular}
  \caption{Performance with the greedy branching rule, data set: n50}
  \label{tab:test.cutmds-n50}
\end{table}

\begin{table}[!htb]
  \centering
  \begin{tabular}[center]{|c|c|c|c|c|c|c|c|}
    \hline
    Num & Dim & Suc & Tim & TD & Nod & UB & Dep \\
    \hline
    16 & 5 & yes & 0.49 & 11 & 307 & 5 & 4 \\
    16 & 5 & yes & 0.78 & 11 & 573 & 5 & 4 \\
    26 & 5 & yes & 6.21 & 14 & 2631 & 8 & 6 \\
    26 & 5 & yes & 9.50 & 14 & 3903 & 8 & 8$(*)$ \\
    36 & 5 & yes & 96.77 & 19 & 23611 & 13 & 12 \\
    36 & 5 & yes & 99.37 & 19 & 25147 & 13 & 12 \\
    46 & 5 & yes & 450.37 & 23 & 48993 & 16 & 16 \\
    46 & 5 & yes & 510.80 & 23 & 67557 & 18 & 15$(*)$ \\
    56 & 5 & outmem &  &  &  & 21 &  \\
    56 & 5 & outmem &&&& 20 &\\
    66 & 5 & outmem &&&& 24 &\\
    66 & 5 & outmem &  &  &  & 25 &  \\
    76 & 5 & outmem &&&& 30 &\\
    76 & 5 & outmem &&&& 29 &\\
    86 & 5 & outmem &&&& 35 &\\
    86 & 5 & outmem &&&& 35 &\\
    \hline
  \end{tabular}
  \caption{Performance with the greedy branching rule, data set: sd5}
  \label{tab:test.cutmds-sd5}
\end{table}

\clearpage
\subsection{Results of the Strong Branching and MDS Cut Generator and Depth First Search}
\label{sec:apd.bac.bran}

\begin{table}[!htb]
  \centering
  \begin{tabular}[center]{|c|c|c|c|c|c|c|c|}
    \hline
    Num & Dim & Suc & Tim & TD & Nod & UB & Dep \\
    \hline
    20 & 5 & yes & 0.24 & 9 & 65 & 4 & 4 \\
    20 & 5 & yes & 0.39 & 9 & 77 & 5 & 5 \\
    30 & 5 & yes & 1.95 & 10 & 249 & 6 & 6 \\
    30 & 5 & yes & 0.90 & 9 & 152 & 4 & 4 \\
    30 & 5 & yes & 4.44 & 13 & 541 & 8 & 8$(*)$ \\
    40 & 5 & yes & 0.74 & 9 & 34 & 5 & 5 \\
    40 & 5 & yes & 22.81 & 15 & 1607 & 10 & 10 \\
    40 & 5 & yes & 45.79 & 16 & 3743 & 11 & 11$(*)$ \\
    50 & 5 & yes & 60.73 & 15 & 3750 & 10 & 10 \\
    50 & 5 & yes & 294.81 & 20 & 18237 & 14 & 14 \\
    50 & 5 & yes & 451.83 & 21 & 22364 & 15 & 15 \\
    60 & 5 & yes & 880.08 & 22 & 33964 & 16 & 16$(*)$ \\
    60 & 5 & outmem &&&& 19 &\\
    60 & 5 & yes & 318.49 & 19 & 11297 & 13 & 13 \\
    70 & 5 & outmem &&&& 21 &\\
    70 & 5 & outmem &&&& 20 &\\
    70 & 5 & outmem &&&& 23 &\\
    80 & 5 & outtime &&&& 20 &\\
    80 & 5 & outmem &&&& 22 &\\
    80 & 5 & outmem &&&& 25 &\\
    90 & 5 & outmem &&&& 29 &\\
    90 & 5 & outmem &&&& 21 &\\
    90 & 5 & outmem &&&& 24 &\\
    \hline
  \end{tabular}
  \caption{Performance with strong branching, data set: d5}
  \label{tab:test.bran-d5}
\end{table}

\begin{table}[!htb]
  \centering
  \begin{tabular}[center]{|c|c|c|c|c|c|c|c|}
    \hline
    Num & Dim & Suc & Tim & TD & Nod & UB & Dep \\
    \hline
    50 & 10 & yes & 68.78 & 16 & 2592 & 5 & 5 \\
    50 & 10 & yes & 6.11 & 13 & 184 & 3 & 3 \\
    50 & 10 & yes & 551.35 & 18 & 21750 & 7 & 7 \\
    60 & 10 & yes & 90.50 & 16 & 2533 & 5 & 5 \\
    60 & 10 & yes & 295.14 & 17 & 7697 & 6 & 6 \\
    60 & 10 & yes & 69.53 & 16 & 2022 & 5 & 4 \\
    70 & 10 & yes & 452.90 & 17 & 7743 & 6 & 6 \\
    70 & 10 & outtime &&&& 10 &\\
    70 & 10 & yes & 1415.18 & 19 & 21407 & 7 & 7 \\
    80 & 10 & outtime &  &  &  & 10 &  \\
    80 & 10 & outtime &  &  &  & 8 &  \\
    80 & 10 & outtime &&&& 15 &\\
    90 & 10 & outtime &&&& 18 &\\
    90 & 10 & outmem &  &  &  & 16 &  \\
    90 & 10 & outtime &&&& 14 &\\
    110 & 10 & outtime &&&& 20 &\\
    120 & 10 & outtime &&&& 22 &\\
    120 & 10 & outtime &&&& 25 &\\
    130 & 10 & outtime &&&& 28 &\\
    \hline
  \end{tabular}
  \caption{Performance with strong branching, data set: d10}
  \label{tab:test.bran-d10}
\end{table}

\begin{table}[!htb]
  \centering
  \begin{tabular}[center]{|c|c|c|c|c|c|c|c|}
    \hline
    Num & Dim & Suc & Tim & TD & Nod & UB & Dep \\
    \hline
    50 & 3 & yes & 18.23 & 19 & 618 & 19 & 19$(*)$ \\
    50 & 3 & yes & 1.87 & 10 & 53 & 12 & 12 \\
    50 & 3 & yes & 28.40 & 20 & 1300 & 18 & 18 \\
    50 & 4 & yes & 93.95 & 20 & 3812 & 16 & 16 \\
    50 & 4 & yes & 20.84 & 16 & 950 & 12 & 12 \\
    50 & 4 & yes & 29.37 & 15 & 1534 & 11 & 11 \\
    50 & 5 & yes & 8.26 & 13 & 279 & 9 & 9 \\
    50 & 5 & yes & 201.13 & 20 & 11376 & 14 & 13 \\
    50 & 5 & yes & 250.46 & 20 & 16958 & 14 & 13 \\
    50 & 6 & yes & 1096.93 & 22 & 60545 & 14 & 14 \\
    50 & 6 & yes & 614.84 & 19 & 45113 & 12 & 11 \\
    50 & 6 & yes & 158.91 & 17 & 7491 & 10 & 10 \\
    50 & 7 & yes & 1525.51 & 22 & 103429 & 13 & 11 \\
    50 & 7 & yes & 35.34 & 14 & 1410 & 7 & 7 \\
    50 & 7 & yes & 173.09 & 16 & 10087 & 8 & 8 \\
    50 & 8 & yes & 3280.20 & 22 & 187925 & 12 & 11 \\
    50 & 8 & yes & 7.80 & 13 & 414 & 4 & 4 \\
    50 & 8 & yes & 29.56 & 14 & 1295 & 5 & 5 \\
    50 & 9 & yes & 117.13 & 16 & 4786 & 6 & 5 \\
    50 & 9 & yes & 2764.54 & 21 & 142913 & 10 & 9 \\
    50 & 9 & yes & 574.78 & 18 & 32367 & 8 & 8 \\
    50 & 10 & yes & 66.55 & 16 & 2443 & 5 & 4 \\
    50 & 10 & yes & 201.70 & 17 & 8021 & 6 & 5 \\
    50 & 10 & yes & 228.01 & 17 & 8064 & 6 & 6 \\
    50 & 11 & yes & 908.46 & 19 & 31971 & 7 & 6 \\
    50 & 11 & yes & 93.19 & 17 & 3541 & 5 & 5 \\
    50 & 11 & yes & 1.27 & 13 & 37 & 2 & 2 \\
    50 & 12 & yes & 1.38 & 14 & 39 & 2 & 2 \\
    50 & 12 & yes & 9.00 & 15 & 252 & 3 & 3 \\
    50 & 12 & yes & 39.67 & 17 & 1223 & 4 & 4 \\
    \hline
  \end{tabular}
  \caption{Performance with strong branching, data set: n50}
  \label{tab:test.bran-n50}
\end{table}

\begin{table}[!htb]
  \centering
  \begin{tabular}[center]{|c|c|c|c|c|c|c|c|}
    \hline
    Num & Dim & Suc & Tim & TD & Nod & UB & Dep \\
    \hline
    16 & 5 & yes & 0.35 & 10 & 130 & 5 & 4 \\
    16 & 5 & yes & 0.44 & 10 & 175 & 5 & 4 \\
    26 & 5 & yes & 5.98 & 14 & 1393 & 8 & 6 \\
    26 & 5 & yes & 8.28 & 15 & 1821 & 8 & 8$(*)$ \\
    36 & 5 & yes & 85.98 & 20 & 12477 & 13 & 12 \\
    36 & 5 & yes & 93.01 & 19 & 11764 & 13 & 12 \\
    46 & 5 & yes & 596.31 & 23 & 42126 & 16 & 16 \\
    46 & 5 & yes & 477.06 & 26 & 36778 & 18 & 15$(*)$ \\
    56 & 5 & outmem &&&& 21 &\\
    56 & 5 & outmem &&&& 20 &\\
    66 & 5 & outmem &&&& 24 &\\
    66 & 5 & outmem &&&& 25 &\\
    76 & 5 & outmem &&&& 30 &\\
    76 & 5 & outmem &&&& 30 &\\
    86 & 5 & outmem &&&& 35 &\\
    86 & 5 & outmem &&&& 37 &\\
    \hline
  \end{tabular}
  \caption{Performance with strong branching, data set: sd5}
  \label{tab:test.bran-sd5}
\end{table}

\clearpage
\subsection{Results of the Strong Branching and MDS Cut Generator and Best First Search}
\label{sec:apd.bac.cutmds}

\begin{table}[!htb]
  \centering
  \begin{tabular}[center]{|c|c|c|c|c|c|c|c|}
    \hline
    Num & Dim & Suc & Tim & TD & Nod & UB & Dep \\
    \hline
    20 & 5 & yes & 0.24 & 9 & 65 & 4 & 4 \\
    20 & 5 & yes & 0.49 & 9 & 96 & 5 & 5 \\
    30 & 5 & yes & 2.10 & 10 & 264 & 6 & 6 \\
    30 & 5 & yes & 0.89 & 9 & 152 & 4 & 4 \\
    30 & 5 & yes & 4.34 & 13 & 513 & 8 & 8$(*)$ \\
    40 & 5 & yes & 0.72 & 8 & 31 & 5 & 5 \\
    40 & 5 & yes & 22.31 & 16 & 1540 & 10 & 10 \\
    40 & 5 & yes & 45.00 & 16 & 3841 & 11 & 11$(*)$ \\
    50 & 5 & yes & 61.19 & 15 & 3754 & 10 & 10 \\
    50 & 5 & yes & 244.03 & 20 & 16455 & 14 & 14 \\
    50 & 5 & yes & 400.39 & 22 & 20545 & 15 & 15 \\
    60 & 5 & outmem &&&& 16 &\\
    60 & 5 & outmem &&&& 19 &\\
    60 & 5 & yes & 306.79 & 18 & 12123 & 13 & 13 \\
    70 & 5 & outmem &&&& 22 &\\
    70 & 5 & outmem &&&& 20 &\\
    70 & 5 & outmem &&&& 23 &\\
    80 & 5 & outmem &&&& 21 &\\
    80 & 5 & outmem &&&& 22 &\\
    80 & 5 & outmem &&&& 25 &\\
    90 & 5 & outmem &&&& 29 &\\
    90 & 5 & outmem &&&& 21 &\\
    90 & 5 & outmem &&&& 24 &\\
    \hline
  \end{tabular}
  \caption{Performance with best first tree search strategy, data set: d5}
  \label{tab:test.sel-d5}
\end{table}

\begin{table}[!htb]
  \centering
  \begin{tabular}[center]{|c|c|c|c|c|c|c|c|}
    \hline
    Num & Dim & Suc & Tim & TD & Nod & UB & Dep \\
    \hline
    50 & 10 & yes & 68.97 & 16 & 2592 & 5 & 5 \\
    50 & 10 & yes & 6.10 & 13 & 184 & 3 & 3 \\
    50 & 10 & yes & 553.80 & 18 & 21750 & 7 & 7 \\
    60 & 10 & yes & 90.75 & 16 & 2533 & 5 & 5 \\
    60 & 10 & yes & 296.55 & 17 & 7697 & 6 & 6 \\
    60 & 10 & yes & 54.45 & 15 & 2380 & 5 & 4 \\
    70 & 10 & yes & 453.98 & 17 & 7743 & 6 & 6 \\
    70 & 10 & outtime &&&& 10 &\\
    70 & 10 & yes & 1416.57 & 19 & 21407 & 7 & 7 \\
    80 & 10 & outtime &&&& 10 &\\
    80 & 10 & outtime &&&& 8 &\\
    80 & 10 & outtime &&&& 15 &\\
    90 & 10 & outtime &&&& 18 &\\
    90 & 10 & outtime &  &  &  & 16 &  \\
    90 & 10 & outtime &&&& 14 &\\
    110 & 10 & outtime &&&& 20 &\\
    120 & 10 & outtime &&&& 22 &\\
    120 & 10 & outtime &&&& 25 &\\
    130 & 10 & outtime &&&& 28 &\\
    \hline
  \end{tabular}
  \caption{Performance with best first tree search strategy, data set: d10}
  \label{tab:test.sel-d10}
\end{table}

\begin{table}[!htb]
  \centering
  \begin{tabular}[center]{|c|c|c|c|c|c|c|c|}
    \hline
    Num & Dim & Suc & Tim & TD & Nod & UB & Dep \\
    \hline
    50 & 3 & yes & 15.49 & 20 & 588 & 19 & 19$(*)$ \\
    50 & 3 & yes & 2.06 & 12 & 69 & 12 & 12 \\
    50 & 3 & yes & 30.43 & 20 & 1411 & 18 & 18 \\
    50 & 4 & yes & 78.82 & 20 & 3469 & 16 & 16 \\
    50 & 4 & no &  &  &  &  &  \\
    50 & 4 & yes & 25.89 & 15 & 1456 & 11 & 11 \\
    50 & 5 & yes & 7.37 & 13 & 261 & 9 & 9 \\
    50 & 5 & no &  &  &  &  &  \\
    50 & 5 & yes & 290.62 & 19 & 23463 & 14 & 13 \\
    50 & 6 & no &  &  &  &  &  \\
    50 & 6 & yes & 519.63 & 19 & 44841 & 12 & 11 \\
    50 & 6 & yes & 147.15 & 16 & 7203 & 10 & 10 \\
    50 & 7 & yes & 1240.54 & 21 & 112366 & 13 & 11 \\
    50 & 7 & yes & 35.02 & 14 & 1411 & 7 & 7 \\
    50 & 7 & yes & 171.64 & 16 & 10087 & 8 & 8 \\
    50 & 8 & yes & 3448.78 & 22 & 250539 & 12 & 11 \\
    50 & 8 & yes & 7.75 & 13 & 414 & 4 & 4 \\
    50 & 8 & yes & 29.25 & 14 & 1295 & 5 & 5 \\
    50 & 9 & yes & 78.47 & 16 & 5030 & 6 & 5 \\
    50 & 9 & yes & 2182.89 & 21 & 142552 & 10 & 9 \\
    50 & 9 & yes & 571.29 & 18 & 32368 & 8 & 8 \\
    50 & 10 & yes & 39.57 & 15 & 2376 & 5 & 4 \\
    50 & 10 & yes & 116.86 & 17 & 7404 & 6 & 5 \\
    50 & 10 & yes & 228.21 & 17 & 8064 & 6 & 6 \\
    50 & 11 & yes & 592.47 & 19 & 31130 & 7 & 6 \\
    50 & 11 & yes & 93.12 & 17 & 3541 & 5 & 5 \\
    50 & 11 & yes & 1.27 & 13 & 37 & 2 & 2 \\
    50 & 12 & yes & 1.38 & 14 & 39 & 2 & 2 \\
    50 & 12 & yes & 9.00 & 15 & 253 & 3 & 3 \\
    50 & 12 & yes & 39.61 & 17 & 1222 & 4 & 4 \\
    \hline
  \end{tabular}
  \caption{Performance with best first tree search strategy, data set: n50}
  \label{tab:test.sel-n50}
\end{table}

\begin{table}[!htb]
  \centering
  \begin{tabular}[center]{|c|c|c|c|c|c|c|c|}
    \hline
    Num & Dim & Suc & Tim & TD & Nod & UB & Dep \\
    \hline
    16 & 5 & yes & 0.34 & 9 & 180 & 5 & 4 \\
    16 & 5 & yes & 0.54 & 10 & 345 & 5 & 4 \\
    26 & 5 & yes & 3.46 & 13 & 1050 & 8 & 6 \\
    26 & 5 & yes & 8.04 & 15 & 1839 & 8 & 8$(*)$ \\
    36 & 5 & yes & 89.81 & 20 & 13844 & 13 & 12 \\
    36 & 5 & yes & 86.77 & 19 & 11768 & 13 & 12 \\
    46 & 5 & no &  &  &  &  &  \\
    46 & 5 & yes & 551.51 & 25 & 51826 & 18 & 15$(*)$ \\
    56 & 5 & outmem &&&& 21 &\\
    56 & 5 & outmem &&&&  &\\
    66 & 5 & outmem &&&& 24 &\\
    66 & 5 & outmem &&&& 25 &\\
    76 & 5 & outmem &&&& 30 &\\
    76 & 5 & outmem &&&& 30 &\\
    86 & 5 & outmem &&&& 35 &\\
    86 & 5 & outmem &&&& 37 &\\
    \hline
  \end{tabular}
  \caption{Performance with best first tree search strategy, data set: sd5}
  \label{tab:test.sel-sd5}
\end{table}

\clearpage
\subsection{Results of the Strong Branching and BIS Cut Generator}
\label{sec:apd.bac.cutbis}

\begin{table}[!htb]
  \centering
  \begin{tabular}[center]{|c|c|c|c|c|c|c|c|}
    \hline
    Num & Dim & Suc & Tim & TD & Nod & UB & Dep \\
    \hline
    20 & 5 & yes & 0.16 & 8 & 63 & 4 & 4 \\
    20 & 5 & yes & 0.44 & 9 & 181 & 5 & 5 \\
    30 & 5 & yes & 1.07 & 10 & 377 & 6 & 6 \\
    30 & 5 & yes & 0.17 & 8 & 57 & 4 & 4 \\
    30 & 5 & yes & 3.83 & 13 & 1447 & 8 & 7 \\
    40 & 5 & yes & 0.55 & 10 & 166 & 5 & 5 \\
    40 & 5 & yes & 14.12 & 15 & 4571 & 10 & 10 \\
    40 & 5 & yes & 22.79 & 16 & 7511 & 11 & 10 \\
    50 & 5 & yes & 15.65 & 15 & 4420 & 10 & 10 \\
    50 & 5 & yes & 89.68 & 20 & 25492 & 14 & 14 \\
    50 & 5 & yes & 129.29 & 20 & 37698 & 15 & 15 \\
    60 & 5 & yes & 177.52 & 22 & 45043 & 16 & 15 \\
    60 & 5 & yes & 366.92 & 25 & 95820 & 19 & 17 \\
    60 & 5 & yes & 72.54 & 18 & 17240 & 13 & 13 \\
    70 & 5 & yes & 1246.22 & 28 & 287853 & 22 & 21 \\
    70 & 5 & yes & 777.33 & 27 & 178746 & 20 & 19 \\
    70 & 5 & yes & 1689.58 & 29 & 393702 & 23 & 23 \\
    80 & 5 & yes & 1221.53 & 28 & 248762 & 21 & 20 \\
    80 & 5 & yes & 990.78 & 28 & 198836 & 22 & 19 \\
    80 & 5 & yes & 2875.70 & 31 & 603491 & 25 & 24 \\
    90 & 5 & outmem &&&& 29 &\\
    90 & 5 & yes & 1395.70 & 28 & 250491 & 21 & 21 \\
    90 & 5 & yes & 2927.97 & 31 & 539796 & 24 & 24 \\
    \hline
  \end{tabular}
  \caption{Performance with BIS cut generator, data set: d5}
  \label{tab:test.cutbis-d5}
\end{table}

\begin{table}[!htb]
  \centering
  \begin{tabular}[center]{|c|c|c|c|c|c|c|c|}
    \hline
    Num & Dim & Suc & Tim & TD & Nod & UB & Dep \\
    \hline
    50 & 10 & yes & 3.97 & 15 & 807 & 5 & 5 \\
    50 & 10 & yes & 0.22 & 12 & 43 & 3 & 3 \\
    50 & 10 & yes & 42.17 & 18 & 9220 & 7 & 7 \\
    60 & 10 & yes & 4.27 & 14 & 753 & 5 & 5 \\
    60 & 10 & yes & 15.43 & 16 & 2722 & 6 & 6 \\
    60 & 10 & yes & 3.50 & 14 & 620 & 5 & 4 \\
    70 & 10 & yes & 17.04 & 16 & 2680 & 6 & 6 \\
    70 & 10 & yes & 796.99 & 21 & 132672 & 10 & 10 \\
    70 & 10 & yes & 52.06 & 19 & 8267 & 7 & 7 \\
    80 & 10 & yes & 903.12 & 21 & 130127 & 10 & 10 \\
    80 & 10 & yes & 164.53 & 21 & 22439 & 8 & 8 \\
    80 & 10 & outtime &&&& 15 &\\
    90 & 10 & outmem &&&& 18 &\\
    90 & 10 & outmem &  &  &  & 16 &  \\
    90 & 10 & outmem &&&& 14 &\\
    110 & 10 & outmem &&&& 20 &\\
    120 & 10 & outmem &&&& 22 &\\
    120 & 10 & outmem &&&& 25 &\\
    130 & 10 & outmem &&&& 28 &\\
    \hline
  \end{tabular}
  \caption{Performance with BIS cut generator, data set: d10}
  \label{tab:test.cutbis-d10}
\end{table}

\begin{table}[!htb]
  \centering
  \begin{tabular}[center]{|c|c|c|c|c|c|c|c|}
    \hline
    Num & Dim & Suc & Tim & TD & Nod & UB & Dep \\
    \hline
    50 & 3 & yes & 20.79 & 22 & 6600 & 19 & 18 \\
    50 & 3 & yes & 4.30 & 15 & 1286 & 12 & 12 \\
    50 & 3 & yes & 17.79 & 21 & 5676 & 18 & 18 \\
    50 & 4 & yes & 50.47 & 21 & 15062 & 16 & 16 \\
    50 & 4 & yes & 14.03 & 16 & 4086 & 12 & 12 \\
    50 & 4 & yes & 9.42 & 15 & 2774 & 11 & 11 \\
    50 & 5 & yes & 9.81 & 14 & 2750 & 9 & 9 \\
    50 & 5 & yes & 85.41 & 20 & 24138 & 14 & 13 \\
    50 & 5 & yes & 72.49 & 19 & 21080 & 14 & 13 \\
    50 & 6 & yes & 268.88 & 21 & 75044 & 14 & 14 \\
    50 & 6 & yes & 86.39 & 19 & 23175 & 12 & 11 \\
    50 & 6 & yes & 37.20 & 16 & 9929 & 10 & 10 \\
    50 & 7 & yes & 333.07 & 21 & 87192 & 13 & 11 \\
    50 & 7 & yes & 9.28 & 14 & 2282 & 7 & 7 \\
    50 & 7 & yes & 21.02 & 15 & 5227 & 8 & 8 \\
    50 & 8 & yes & 447.77 & 21 & 112751 & 12 & 11 \\
    50 & 8 & yes & 0.57 & 11 & 123 & 4 & 4 \\
    50 & 8 & yes & 1.99 & 12 & 457 & 5 & 5 \\
    50 & 9 & yes & 7.96 & 15 & 1700 & 6 & 5 \\
    50 & 9 & yes & 324.54 & 20 & 77184 & 10 & 9 \\
    50 & 9 & yes & 66.37 & 18 & 15087 & 8 & 8 \\
    50 & 10 & yes & 3.96 & 15 & 801 & 5 & 4 \\
    50 & 10 & yes & 8.39 & 16 & 1774 & 6 & 5 \\
    50 & 10 & yes & 14.29 & 16 & 2996 & 6 & 6 \\
    50 & 11 & yes & 54.82 & 18 & 11364 & 7 & 6 \\
    50 & 11 & yes & 5.75 & 16 & 1078 & 5 & 5 \\
    50 & 11 & yes & 0.03 & 1 & 2 & 2 & 2 \\
    50 & 12 & yes & 0.05 & 5 & 7 & 2 & 2 \\
    50 & 12 & yes & 0.34 & 14 & 59 & 3 & 3 \\
    50 & 12 & yes & 1.64 & 15 & 293 & 4 & 4 \\
    \hline
  \end{tabular}
  \caption{Performance with BIS cut generator, data set: n50}
  \label{tab:test.cutbis-n50}
\end{table}

\begin{table}[!htb]
  \centering
  \begin{tabular}[center]{|c|c|c|c|c|c|c|c|}
    \hline
    Num & Dim & Suc & Tim & TD & Nod & UB & Dep \\
    \hline
    16 & 5 & yes & 0.18 & 10 & 78 & 5 & 4 \\
    16 & 5 & yes & 0.23 & 11 & 108 & 5 & 4 \\
    26 & 5 & yes & 3.37 & 14 & 1356 & 8 & 6 \\
    26 & 5 & yes & 2.50 & 14 & 1016 & 8 & 7 \\
    36 & 5 & yes & 38.78 & 18 & 12909 & 13 & 12 \\
    36 & 5 & yes & 37.01 & 21 & 13183 & 13 & 12 \\
    46 & 5 & yes & 188.91 & 23 & 59063 & 16 & 16 \\
    46 & 5 & yes & 139.06 & 25 & 43252 & 18 & 14 \\
    56 & 5 & yes & 727.58 & 21 & 202080 & 21 & 19 \\
    56 & 5 & yes & 882.08 & 29 & 242383 & 24 & 20 \\
    66 & 5 & yes & 2126.74 & 31 & 536278 & 24 & 22 \\
    66 & 5 & yes & 2570.02 & 32 & 648729 & 25 & 24 \\
    76 & 5 & outmem &&&& 30 &\\
    76 & 5 & outmem &&&& 29 &\\
    86 & 5 & outmem &&&& 35 &\\
    86 & 5 & outmem &&&& 35 &\\
    \hline
  \end{tabular}
  \caption{Performance with BIS cut generator, data set: sd5}
  \label{tab:test.cutbis-sd5}
\end{table}


\clearpage
\section{Results of the Binary Search Algorithm}
\label{sec:apd.bin}

\begin{table}[!htb]
  \centering
  \begin{tabular}[center]{|c|c|c|c|c|}
    \hline
    Num & Dim & Suc & Tim & Dep \\
    \hline
    20 & 5 & yes & 0.08 & 4 \\
    20 & 5 & yes & 0.27 & 5 \\
    30 & 5 & yes & 0.65 & 6 \\
    30 & 5 & yes & 0.09 & 4 \\
    30 & 5 & yes & 3.30 & 7 \\
    40 & 5 & yes & 0.35 & 5 \\
    40 & 5 & yes & 18.93 & 10 \\
    40 & 5 & yes & 27.71 & 10 \\
    50 & 5 & yes & 24.27 & 10 \\
    50 & 5 & yes & 226.86 & 14 \\
    50 & 5 & yes & 349.705 & 15 \\
    60 & 5 & yes & 348.43 & 15 \\
    60 & 5 & yes & 2389.03 & 17 \\
    60 & 5 & yes & 121.78 & 13 \\
    70 & 5 & outtime &&\\
    70 & 5 & outtime &&\\
    70 & 5 & outtime &&\\
    80 & 5 & outtime &&\\
    80 & 5 & outtime &&\\
    80 & 5 & outtime &&\\
    90 & 5 & outtime &&\\
    90 & 5 & outtime &&\\
    90 & 5 & outtime &&\\
    \hline
  \end{tabular}
  \caption{Performance of the binary search, data set: d5}
  \label{tab:test.bin-d5}
\end{table}

\begin{table}[!htb]
  \centering
  \begin{tabular}[center]{|c|c|c|c|c|}
    \hline
    Num & Dim & Suc & Tim & Dep \\
    \hline
    50 & 10 & yes & 1.45 & 5 \\
    50 & 10 & yes & 0.09 & 3 \\
    50 & 10 & yes & 19.13 & 7 \\
    60 & 10 & yes & 1.68 & 5 \\
    60 & 10 & yes & 6.22 & 6 \\
    60 & 10 & yes & 1.21 & 4 \\
    70 & 10 & yes & 7.05 & 6 \\
    70 & 10 & yes & 855.21 & 10 \\
    70 & 10 & yes & 23.84 & 7 \\
    80 & 10 & yes & 1062.49 & 10 \\
    80 & 10 & yes & 103.47 & 8 \\
    80 & 10 & outtime &&\\
    90 & 10 & outtime &&\\
    90 & 10 & outtime &&\\
    90 & 10 & outtime &&\\
    110 & 10 & outtime &&\\
    120 & 10 & outtime &&\\
    120 & 10 & outtime &&\\
    130 & 10 & outtime &&\\
    \hline
  \end{tabular}
  \caption{Performance of the binary search, data set: d10}
  \label{tab:test.bin-d10}
\end{table}

\begin{table}[!htb]
  \centering
  \begin{tabular}[center]{|c|c|c|c|c|}
    \hline
    Num & Dim & Suc & Tim & Dep \\
    \hline
    50 & 3 & yes & 48.63 & 18 \\
    50 & 3 & yes & 8.10 & 12 \\
    50 & 3 & yes & 71.81 & 18 \\
    50 & 4 & yes & 217.37 & 16 \\
    50 & 4 & yes & 20.78 & 12 \\
    50 & 4 & yes & 15.93 & 11 \\
    50 & 5 & yes & 12.13 & 9 \\
    50 & 5 & yes & 135.55 & 13 \\
    50 & 5 & yes & 164.87 & 13 \\
    50 & 6 & yes & 509.81 & 14 \\
    50 & 6 & yes & 171.37 & 11 \\
    50 & 6 & yes & 54.10 & 10 \\
    50 & 7 & yes & 448.17 & 11 \\
    50 & 7 & yes & 5.38 & 7 \\
    50 & 7 & yes & 18.76 & 8 \\
    50 & 8 & yes & 451.09 & 11 \\
    50 & 8 & yes & 0.22 & 4 \\
    50 & 8 & yes & 0.99 & 5 \\
    50 & 9 & yes & 3.95 & 5 \\
    50 & 9 & yes & 213.69 & 9 \\
    50 & 9 & yes & 48.82 & 8 \\
    50 & 10 & yes & 1.70 & 4 \\
    50 & 10 & yes & 2.54 & 5 \\
    50 & 10 & yes & 5.54 & 6 \\
    50 & 11 & yes & 32.68 & 6 \\
    50 & 11 & yes & 1.94 & 5 \\
    50 & 11 & yes & 0.01 & 2 \\
    50 & 12 & yes & 0.01 & 2 \\
    50 & 12 & yes & 0.12 & 3 \\
    50 & 12 & yes & 0.52 & 4 \\
    \hline
  \end{tabular}
  \caption{Performance of the binary search, data set: n50}
  \label{tab:test.bin-n50}
\end{table}

\begin{table}[!htb]
  \centering
  \begin{tabular}[center]{|c|c|c|c|c|}
    \hline
    Num & Dim & Suc & Tim & Dep \\
    \hline
    16 & 5 & yes & 0.16 & 4 \\
    16 & 5 & yes & 0.15 & 4 \\
    26 & 5 & yes & 1.72 & 6 \\
    26 & 5 & yes & 1.97 & 7 \\
    36 & 5 & yes & 116.26 & 12 \\
    36 & 5 & yes & 99.88 & 12 \\
    46 & 5 & yes & 960.48 & 16 \\
    46 & 5 & yes & 457.56 & 14 \\
    56 & 5 & yes & 3585.18 & 19 \\
    56 & 5 & outtime &&\\
    66 & 5 & outtime &&\\
    66 & 5 & outtime &  &  \\
    76 & 5 & outtime &&\\
    76 & 5 & outtime &&\\
    86 & 5 & outtime &&\\
    86 & 5 & outtime &&\\
    \hline
  \end{tabular}
  \caption{Performance of the binary search, data set: sd5}
  \label{tab:test.bin-sd5}
\end{table}

\clearpage
\section{Results of the Primal-Dual Algorithm}
\label{sec:apd.pd}

\begin{table}[!htb]
  \centering
  \begin{tabular}[center]{|c|c|c|c|c|}
    \hline
    Num & Dim & Suc & Tim & Dep \\
    \hline
    20 & 5 & yes & 0.09 & 4 \\
    20 & 5 & yes & 0.90 & 5 \\
    30 & 5 & yes & 1.95 & 6 \\
    30 & 5 & yes & 0.27 & 4 \\
    30 & 5 & yes & 5.0 & 7 \\
    40 & 5 & yes & 1.19 & 5 \\
    40 & 5 & yes & 144.99 & 10 \\
    40 & 5 & yes & 52.79 & 10 \\
    50 & 5 & yes & 35.66 & 10 \\
    50 & 5 & yes & 1030.45 & 14 \\
    50 & 5 & yes & 5465.40 & 15 \\
    60 & 5 & yes & 795.975 & 15 \\
    60 & 5 & yes & 8810.34 & 17 \\
    60 & 5 & yes & 99.21 & 13 \\
    70 & 5 & outtime &&\\
    70 & 5 & yes & 26274.70 & 19 \\
    70 & 5 & outtime &&\\
    80 & 5 & yes & 13232.58 & 20 \\
    80 & 5 & yes & 3899.48 & 19 \\
    80 & 5 & outtime &&\\
    90 & 5 & outtime &&\\
    90 & 5 & yes & 12821.73 & 21 \\
    90 & 5 & outtime &&\\
    \hline
  \end{tabular}
  \caption{Performance of the primal-dual algorithm, data set: d5}
  \label{tab:test.pd-d5}
\end{table}

\begin{table}[!htb]
  \centering
  \begin{tabular}[center]{|c|c|c|c|c|}
    \hline
    Num & Dim & Suc & Tim & Dep \\
    \hline
    50 & 10 & yes & 45.05 & 5 \\
    50 & 10 & yes & 2.41 & 3 \\
    50 & 10 & yes & 231.06 & 7 \\
    60 & 10 & yes & 16.16 & 5 \\
    60 & 10 & yes & 35.98 & 6 \\
    60 & 10 & yes & 5.50 & 4 \\
    70 & 10 & yes & 26.22 & 6 \\
    70 & 10 & outtime &&\\
    70 & 10 & yes & 40.88 & 7 \\
    80 & 10 & yes & 38915.29 & 10 \\
    80 & 10 & yes & 121.22 & 8 \\
    80 & 10 & outtime &&\\
    90 & 10 & outtime &&\\
    90 & 10 & outtime &&\\
    90 & 10 & outtime &&\\
    110 & 10 & outtime &&\\
    120 & 10 & outtime & &\\
    120 & 10 & outtime & &\\
    130 & 10 & outtime & &\\
    \hline
  \end{tabular}
  \caption{Performance of the primal-dual algorithm, data set: d10}
  \label{tab:test.pd-d10}
\end{table}

\begin{table}[!htb]
  \centering
  \begin{tabular}[center]{|c|c|c|c|c|}
    \hline
    Num & Dim & Suc & Tim & Dep \\
    \hline
    50 & 3 & yes & 47.94 & 18 \\
    50 & 3 & yes & 1.30 & 12 \\
    50 & 3 & yes & 60.31 & 18 \\
    50 & 4 & yes & 287.33 & 16 \\
    50 & 4 & yes & 16.07 & 12 \\
    50 & 4 & yes & 25.47 & 11 \\
    50 & 5 & yes & 9.53 & 9 \\
    50 & 5 & yes & 278.84 & 13 \\
    50 & 5 & yes & 514.03 & 13 \\
    50 & 6 & yes & 16052.87 & 14 \\
    50 & 6 & yes & 1031.67 & 11 \\
    50 & 6 & yes & 279.75 & 10 \\
    50 & 7 & yes & 4797.68 & 11 \\
    50 & 7 & yes & 12.94 & 7 \\
    50 & 7 & yes & 41.91 & 8 \\
    50 & 8 & yes & 16336.03 & 11 \\
    50 & 8 & yes & 1.33 & 4 \\
    50 & 8 & yes & 4.71 & 5 \\
    50 & 9 & yes & 4.23 & 5 \\
    50 & 9 & yes & 10425.83 & 9 \\
    50 & 9 & yes & 2440.02 & 8 \\
    50 & 10 & yes & 8.03 & 4 \\
    50 & 10 & yes & 7.82 & 5 \\
    50 & 10 & yes & 112.04 & 6 \\
    50 & 11 & yes & 171.47 & 6 \\
    50 & 11 & yes & 14.90 & 5 \\
    50 & 11 & yes & 2.77 & 2 \\
    50 & 12 & yes & 1.80 & 2 \\
    50 & 12 & yes & 2.41 & 3 \\
    50 & 12 & yes & 4.28 & 4 \\
    \hline
  \end{tabular}
  \caption{Performance of the primal-dual algorithm, data set: n50}
  \label{tab:test.pd-n50}
\end{table}

\begin{table}[!htb]
  \centering
  \begin{tabular}[center]{|c|c|c|c|c|}
    \hline
    Num & Dim & Suc & Tim & Dep \\
    \hline
    16 & 5 & yes & 0.32 & 4 \\
    16 & 5 & yes & 0.24 & 4 \\
    26 & 5 & yes & 3.48 & 6 \\
    26 & 5 & yes & 5.08 & 7 \\
    36 & 5 & yes & 394.41 & 12 \\
    36 & 5 & yes & 715.92 & 12 \\
    46 & 5 & yes & 23908.49 & 16 \\
    46 & 5 & yes & 4093.35 & 14 \\
    56 & 5 & outtime & &\\
    56 & 5 & outtime & &\\
    66 & 5 & outtime & &\\
    66 & 5 & outtime & &\\
    76 & 5 & outtime & &\\
    76 & 5 & outtime & &\\
    86 & 5 & outtime & &\\
    86 & 5 & outtime & &\\
    \hline
  \end{tabular}
  \caption{Performance of the primal-dual algorithm, data set: sd5}
  \label{tab:test.pd-sd5}
\end{table}

\doublespace
\chapter*{Curriculum Vita}

\pagestyle{empty}
\thispagestyle{empty}
\addtocontents{toc}{\contentsline {chapter}{Curriculum Vita}{}}

\section*{Name}

\begin{flushleft}
  Dan Chen
\end{flushleft}

\section*{University Attended}

\begin{flushleft}
  Master of Computer Science, University of New Brunswick, 2007

  B.S., Computer Science, Liaoning University, Shenyang, Liaoning, China, 2003
\end{flushleft}

\section*{Publications}

\begin{flushleft}
  D. Bremner, D. Chen, J. Iacono, S. Langerman, and P. Morin, \textit{Output-sensitive algorithms for Tukey depth and related problems}, Submitted, 2006.
\end{flushleft}

\section*{Conference Presentations}

\begin{flushleft}
  D. Bremner and D. Chen (speaker), \textit{A branch and cut algorithm for the halfspace depth problem}, Poster presentation, CAIMS-MITACS Annual Conference, Toronto, Canada. June 2006.

  D. Bremner and D. Chen (speaker), \textit{A branch and cut algorithm for the halfspace depth problem}, Computer Science Seminar Series, University of New Brunswick. July, 2006.

  D. Bremner and D. Chen (speaker), \textit{A branch and cut algorithm for the halfspace depth problem}, Radcliffe Institute Seminar on Computational Aspects of Statistical Data Depth Analysis, Cambridge, MA, USA. July, 2006.
\end{flushleft}

\end{document}